\documentclass[a4paper,11pt]{article}
\usepackage{aaskaiid}
\usepackage{xcolor}
\usepackage{siunitx}
\usepackage{physics}
\usepackage{orcidlink}
\newcommand{\grammage}[1]{g cm$^{-2}$}
\newcommand{\xmax}[1]{$X_\text{max}$}
\newcommand{\erad}[1]{$E_\text{rad}$}
\newcommand{\vxvxb}[1]{$\vec{v} \times ( \vec{v} \times \vec{B} )$} 
\newcommand{\vxb}[1]{$\vec{v} \times \times \vec{B}$}

\graphicspath{figures}

\title{Interferometric Analysis of Air-shower Radio Emission in the Near Field with an Information Field Theory Approach}
\ShortTitle{Air Shower Interferometry and Information Field Theory}

\author[a]{Keito~Watanabe\orcidlink{0000-0003-0599-4035}}
\author[b]{Karen~Terveer\orcidlink{0009-0002-9594-0419}}
\emailAdd{keito.watanabe@kit.edu,karen.terveer@fau.de}

\author[b]{Sjoerd~Bouma\orcidlink{0000-0002-6959-2302}}
\author[c]{Justin~Bray\orcidlink{0000-0002-0963-0223}}
\author[d,e]{Stijn~Buitink\orcidlink{0000-0002-6177-497X}}
\author[d,e]{Arthur~Corstanje\orcidlink{0000-0001-5992-6228}}
\author[d]{Vital~De Henau\orcidlink{0009-0003-0337-3558}}
\author[a,d]{Tim~Huege\orcidlink{0000-0002-2783-4772}}
\author[f]{Edwin~Dickinson\orcidlink{0000-0003-0834-4708}}
\author[g,h]{Vincent~Eberle\orcidlink{0000-0002-5713-3475}} 
\author[g,h,i]{Torsten~En{\ss}lin\orcidlink{0000-0001-5246-1624}} 
\author[j,k]{Brian~Hare\orcidlink{0000-0001-5138-1235}}
\author[l,v]{Haoning~He\orcidlink{0000-0002-8941-9603}}
\author[e,d,m]{J\"org~H\"orandel\orcidlink{0000-0001-6604-547X}}

\author[f]{Clancy~James\orcidlink{0000-0002-6437-6176}}
\author[b]{Philipp~Laub\orcidlink{0009-0003-2617-9109}}
\author[l]{Xingyu Li\orcidlink{0009-0003-1209-2643}}
\author[a]{Hermann-Josef~Mathes}
\author[e,m]{Katharine~Mulrey\orcidlink{0000-0001-8026-8020}}
\author[b,n]{Anna~Nelles\orcidlink{0000-0002-1720-6350}}
\author[a]{Subhadip~Saha\orcidlink{0000-0003-2435-8317}}
\author[w]{Felix~Schl\"uter\orcidlink{0000-0002-5545-4363}}

\author[j]{Olaf~Scholten\orcidlink{0000-0003-3649-1254}}
\author[c]{Ralph~Spencer\orcidlink{0009-0009-6015-1787}}
\author[k]{Christopher~Sterpka\orcidlink{0000-0001-8217-0836}}

\author[p]{Satyendra~Thoudam\orcidlink{0000-0002-7066-3614}}
\author[q]{Gia~Trinh\orcidlink{0000-0002-5352-5092}}
\author[k]{Paulina~Turekova\orcidlink{0009-0006-1262-7507}}
\author[a]{Darko~Veberic\orcidlink{0000-0003-2683-1526}}

\author[s,t]{Chao~Zhang\orcidlink{0000-0001-9366-0056}}
\author[u]{Pengfei~Zhang\orcidlink{0000-0002-6855-5315}}
\author[l]{Yi~Zhang\orcidlink{0000-0001-6223-4724}}

\ShortName{K.~Watanabe, K.~Terveer et al.} 

\newcommand{\affilASTRON}{Netherlands Institute for Radio Astronomy (ASTRON), Dwingeloo, The Netherlands}
\newcommand{\affilCanTho}{Physics Education Department, School of Education, Can Tho University, Campus~II, 3/2 Street, Ninh Kieu District, Can Tho City, Viet Nam}
\newcommand{\affilCurtin}{International Centre for Radio Astronomy Research, Curtin University, Bentley, 6102, WA, Australia}
\newcommand{\affilDESY}{Deutsches Elektronen-Synchrotron DESY, Platanenallee~6, 15738 Zeuthen, Germany}
\newcommand{\affilErlangen}{Erlangen Centre for Astroparticle Physics, Friedrich-Alexander-Universit\"at Erlangen-N\"urnberg, 91058 Erlangen, Germany}
\newcommand{\affilGorlitz}{Deutsches Zentrum f\"ur Astrophysik, Postplatz~1, 02826 Görlitz, Germany}
\newcommand{\affilGroningen}{Kapteyn Astronomical Institute, University of Groningen, P.O.~Box 72, 9700 AB Groningen, Netherlands}
\newcommand{\affilHefei}{School of Astronomy and Space Science, University of Science and Technology of China, Hefei 230026, China}
\newcommand{\affilKanpur}{Department of Physics, Indian Institute of Technology Kanpur, Kanpur, UP-208016, India}
\newcommand{\affilKeyNanjing}{Key Laboratory of Modern Astronomy and Astrophysics, Nanjing University, Ministry of Education, Nanjing 210023, China}
\newcommand{\affilKIT}{Institut f\"ur Astroteilchenphysik, Karlsruhe Institute of Technology (KIT), P.O.~Box 3640, 76021 Karlsruhe, Germany}
\newcommand{\affilKhalifa}{Department of Physics, Khalifa University, P.O.~Box 127788, Abu Dhabi, United Arab Emirates}
\newcommand{\affilManchester}{Jodrell Bank Centre for Astrophysics, Department of Physics and Astronomy, University of Manchester, Manchester M13 9PL, UK}
\newcommand{\affilMaxPlanck}{Max-Planck Institut f\"ur Astrophysik, Karl-Schwarzschild-Str.~1, 85748 Garching, Germany}
\newcommand{\affilMunich}{Ludwig-Maximilians-Universit\"at M\"unchen (LMU), Geschwister-Scholl-Platz~1, 80539 M\"unchen, Germany}
\newcommand{\affilNanjing}{School of Astronomy and Space Science, Nanjing University, Nanjing 210023, China}
\newcommand{\affilNijmegen}{Department of Astrophysics/IMAPP, Radboud University Nijmegen, P.O.~Box 9010, 6500 GL Nijmegen, The Netherlands}
\newcommand{\affilNikhef}{Nikhef, Science Park Amsterdam, 1098 XG Amsterdam, The Netherlands}
\newcommand{\affilPurpleMt}{Key Laboratory of Dark Matter and Space Astronomy, Purple Mountain Observatory, Chinese Academy of Sciences, No.~10 Yuanhua Road, Nanjing, China}
\newcommand{\affilULB}{Universit\'e Libre de Bruxelles, Science Faculty CP230, B-1050 Brussels, Belgium}
\newcommand{\affilVUB}{Inter-University Institute For High Energies (IIHE), Vrije Universiteit Brussel (VUB), Pleinlaan 2, 1050 Brussels, Belgium}
\newcommand{\affilXidian}{School of Electronic Engineering, Xidian University, No.2 South Taibai Road, Xi'an, China}
\newcommand{\affilSKA}{SKA Observatory, Jodrell Bank, Lower Withington, Macclesfield, SK11 9FT, UK}

\affiliation[a]{\affilKIT}
\affiliation[b]{\affilErlangen}
\affiliation[c]{\affilManchester}
\affiliation[d]{\affilVUB}
\affiliation[e]{\affilNijmegen}
\affiliation[f]{\affilCurtin}
\affiliation[g]{\affilMaxPlanck}
\affiliation[h]{\affilMunich}
\affiliation[i]{\affilGorlitz}
\affiliation[j]{\affilGroningen}
\affiliation[k]{\affilASTRON}
\affiliation[l]{\affilPurpleMt}
\affiliation[m]{\affilNikhef}
\affiliation[n]{\affilDESY}
\affiliation[o]{\affilKanpur}
\affiliation[p]{\affilKhalifa}
\affiliation[q]{\affilCanTho}
\affiliation[r]{\affilSKA}
\affiliation[s]{\affilNanjing}
\affiliation[t]{\affilKeyNanjing}
\affiliation[u]{\affilXidian}
\affiliation[v]{\affilHefei}
\affiliation[w]{\affilULB}

\abstract{
    Current reconstruction techniques for air-shower radio emission generated by cosmic rays have shown great success, having been applied to several radio detectors over the last decade. Nevertheless, they are limited by their high computational cost, simplified approximations, and signal information used for reconstruction. As such, advanced analyses are required to not only be able to perform a holistic reconstruction of all parameters, but also to conduct near-field interferometry of the air shower. This can be achieved through Information Field Theory (IFT), an imaging reconstruction framework based on Bayesian inference that can extract all available information within the signal to infer distributions of field-like quantities. In this chapter, we highlight current novel approaches that use IFT for air shower reconstruction, and the potential of their applicability towards SKA-Low.  
}

\begin{document}
\maketitle


\section{Introduction}
\label{sec:CRDetection}

Recently, the progress of radio detection of cosmic rays has grown significantly. Since the 1960s, electrons and positrons produced within the electromagnetic component of the air shower, induced by high-energy cosmic rays interacting with our atmosphere, have been shown to generate radio waves at MHz frequencies. The dominant emission mechanism is called geomagnetic emission, where radio waves are generated when positrons and electrons experience deflections from the geomagnetic field. The secondary effect, sub-dominant in air, is the charge-excess emission, where the continuous build-up of electrons in the shower-front stemming from atmospheric molecules throughout the longitudinal evolution of the shower generates a net charge-excess. The combined emission from the air shower can be readily observed through a multi-array of digital radio antennas, where each antenna measures a single, non-repeating nanosecond pulse \citep{huege_radio_2016}. With its high duty cycle and inexpensive deployment costs, radio detectors have not only been utilized to perform hybrid measurements in conjunction with current particle detectors such as the Pierre Auger Observatory \citep{AbdulHalim:20258h}, but radio signals from air showers have also been measured through existing radio telescopes such as the LOw Frequency ARray (LOFAR, \citealp{van_haarlem_lofar_2013, Terveer:2024pnt}), which uses particle detectors as an external trigger for cosmic ray detection. \par

The measurements of the characteristics of cosmic rays through radio detection are currently performed through simulation-based reconstructions. Here, typically the voltage-based fluence, i.e. the time integral of the squared voltage trace within a signal window, is used for the reconstruction. This quantity, which only varies as a function of the antenna position, is compared through $\chi^2$-minimization against several outcomes generated from simulation frameworks. Among these are CoREAS \citep{Huege:2013vt} and ZHAires \citep{alvarez_zhaires_2012}, which simulate the radio waves generated from individual electrons and positrons with Monte Carlo algorithms, as well as MGMR3D \citep{Mitra:2023zjf}, which semi-analytically calculates the macroscopic radio emission from parameterizations of the charge distribution. Through this method, current radio observatories such as LOFAR and the Auger Engineering Radio Array (AERA) have measured both the cosmic ray energy and the atmospheric depth at shower maximum, $X_\mathrm{max}$, with high resolution, independently without explicitly relying on measurements from particle-based detectors \citep{corstanje_depth_2021, PierreAuger:2023rgk}. Reconstructing $X_\mathrm{max}$ precisely, in particular, is crucial, as it contains valuable insights about the mass composition, i.e. the particle type, of cosmic rays. In a simulation study, this method has also been applied to the AA$^*$ layout of SKA-Low, showcasing an $X_\mathrm{max}$ resolution of $< 10$ \grammage, at least a factor of 2 lower than the resolution obtained from LOFAR and AERA \citep{corstanje_lofar-style_2025, Corstanje01.2026.SKA}. \par

However, this approach has limitations. Firstly, more than twenty simulations need to be performed per measured air shower, which, due to the Monte-Carlo nature of the simulations, can take up to one CPU day per simulation. Secondly, a simultaneous fit for all cosmic ray parameters (energy, $X_\mathrm{max}$, arrival direction) is not possible with this method. The radio emission also depends on more than these three parameters, namely characteristics that contain additional information about the mass composition of the cosmic ray, that cannot be fully reconstructed with this current method \citep{Corstanje:2023uyg, Corstanje01.2026.SKA}. \par 

\begin{figure}
    \centering
	\includegraphics[width=0.495\columnwidth]{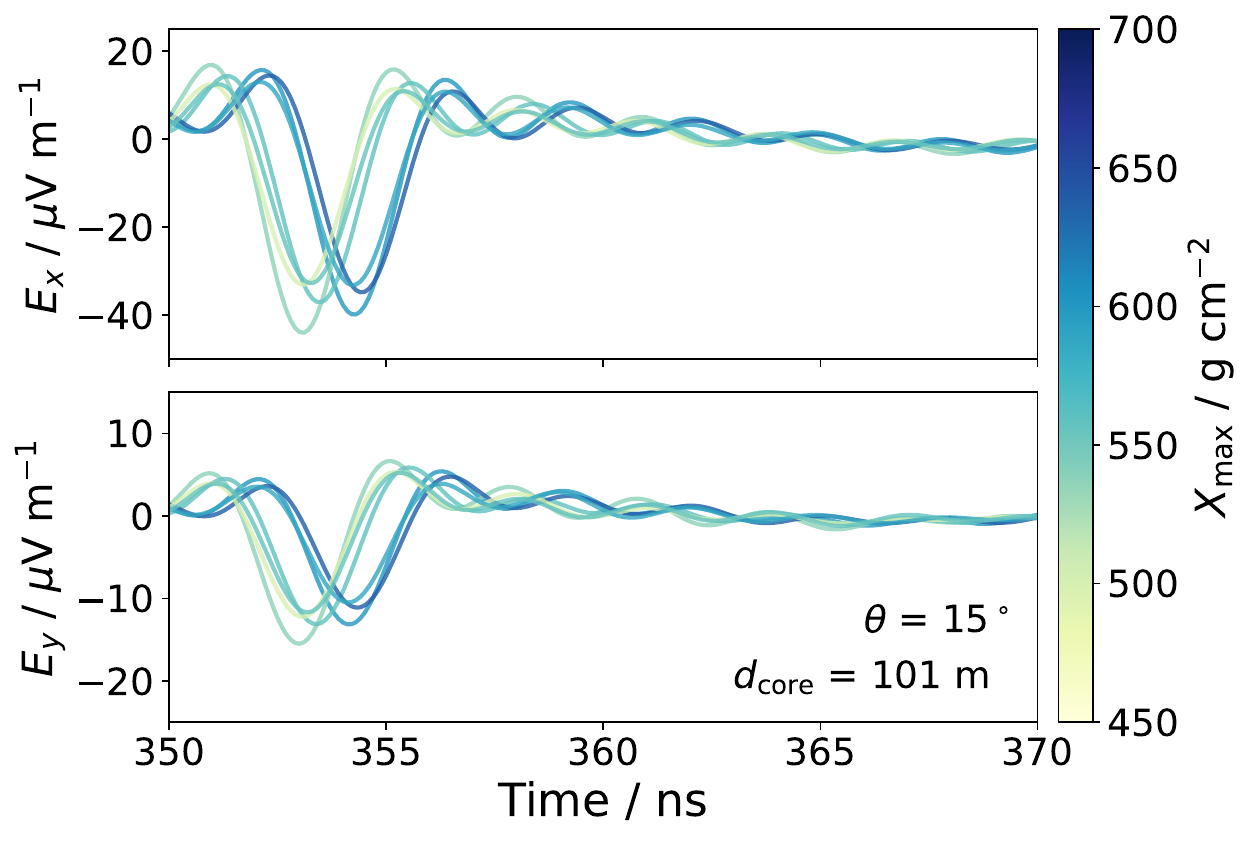}
    \includegraphics[width=0.495\columnwidth]{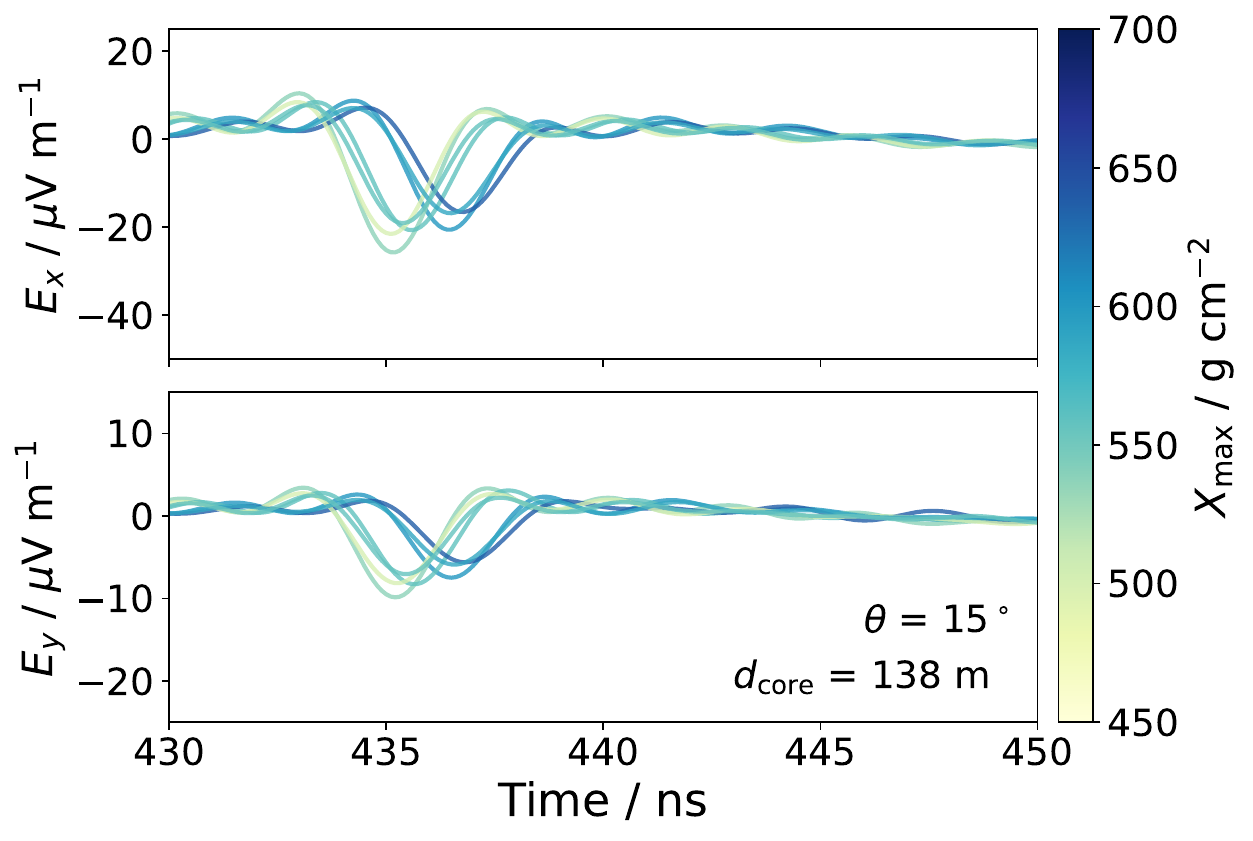}
    \caption{Example of simulated electric field traces from a cosmic ray with primary energy $E = 10^{16}$ eV and zenith angle of 15$^\circ$ at an antenna $d_\mathrm{core} = 101$ m (left) and 138 m (right) from the shower core. The traces are shown for varying values of $X_\mathrm{max}$, indicated by the colorbar. The top and bottom figure show the $x$ and $y$ polarizations of the electric field trace, respectively. Traces are band-pass filtered to [50, 350] MHz to emulate the available bandwidth for SKA-Low. }
    \label{fig:example_traces}
\end{figure}

Lastly, this approach only uses the signal fluence for the reconstruction. To ensure that the method remains bias-free, many events are cut through a strict selection criterion, which reduces the total number of usable cosmic rays. Furthermore, the cosmic ray signal contains additional information, such as its timing and pulse shape, which is not used with this approach. Fig.~\ref{fig:example_traces} shows an example of electric field traces with varying values of $X_\mathrm{max}$ and antenna position, showcasing the wealth of information that can be used in addition to the total energy fluence in the pulses. \par 

The time-varying charged currents of the radio emission can be described with a charged distribution that behaves as a thin, pancake-like structure. This structure flattens out rapidly as a function of time, within $\sim \mu s$ \citep{scholten_analytic_2018}. As such, far-field approximations of the radio emission cannot accurately describe the structure of the air shower profile. Indeed, interferometric techniques that assume a coherent emission at each point in the atmosphere have been performed with idealized antenna layouts \citep{schluter_expected_2021, schoorlemmer_radio_2021} as well as with the AA$^*$ layout of SKA-Low \citep{Corstanje01.2026.SKA}, however they are currently not able to recover the time-varying structure of the air shower profile. Studies have shown that this can be performed by convolving the observed electric fields (including considerations of the antenna geometry) with the expected emission distribution \citep{Scholten:2024upn}. \par 

To circumvent and improve on these various limitations, we are now exploring novel methods to reconstruct the cosmic ray properties through the use of information field theory \citep[IFT,][]{ensslin2019information}. This approach utilizes Bayesian reasoning to not only include prior information based on our physical understanding of air shower physics, but also to infer different realizations of the possible parameters that can produce the observed data, while encompassing the detector response and noise sources within the framework (see Section \ref{sec:IFT}). \par 

Such an improved reconstruction is highly relevant for cosmic ray observations with SKA-Low. Due to its vast number of radio antennas, its large frequency bandwidth and low noise amplitude, we expect to observe the signal from the radio emission of air showers in SKA-Low with unprecedented accuracy. Its large number of antennas is well-suited for an interferometric approach, and the increased frequency bandwidth up to 350 MHz allows more signal to be captured. The enlarged bandwidth is especially crucial for reconstruction of anomalous showers that can aid in understanding the hadronic interactions involved within the air shower \citep{de_henau_investigating_2025, Buitink01.2026.SKA}. The potential for SKA-Low to reconstruct other mass-sensitive parameters has already been studied through a likelihood-based approach \citep{Corstanje:2023uyg, Corstanje01.2026.SKA}. 

In this chapter, we showcase current approaches that utilize IFT in the context of reconstructing cosmic ray air-showers with dense radio antenna arrays. The first approach, highlighted in Section \ref{sec:HolisticRecoLOFAR}, uses the energy fluence and signal timing from all antennas within the LOFAR antenna layout to simultaneously reconstruct all relevant cosmic ray parameters. The second approach, shown in Section \ref{sec:ProfileRecoTS}, focuses on a near-field interferometric approach to reconstruct the entire longitudinal profile of the air shower within an idealized star-shaped antenna layout. In Section \ref{sec:Outlook}, we discuss the application of these methods towards the AA$^*$ layout of SKA-Low. \par

\section{Information Field Theory}
\label{sec:IFT}
Information Field Theory (IFT) provides a framework for signal reconstruction by treating physical fields as continuous mathematical objects governed by probability distributions \citep{ensslin2019information}. This probabilistic approach makes it a powerful tool for analyzing complex, high-dimensional data.

\par At its core, IFT applies Bayesian reasoning to infer an unknown signal field, $\phi$, from observed data, $d$. This is achieved by Bayes' theorem, where the posterior probability $P(\phi|d)$ is proportional to the likelihood $P(d|\phi)$ and the prior $P(\phi)$. The likelihood term models the entire measurement process, from the instrument's response to its noise properties, while the prior incorporates existing physical knowledge about the signal, such as its expected correlation structure. In many cases, these relationships can be sufficiently described with Gaussian distributions:
\begin{align}
    P(d|\phi) &\propto \exp\left(-\frac{1}{2}(d - R(\phi))^\dagger N^{-1} (d - R(\phi))\right) \\
    P(\phi) &\propto \exp\left(-\frac{1}{2}\phi^\dagger S^{-1} \phi\right)
    \label{eq:ift_likeli_prior}
\end{align}
Here, $R(\phi)$ represents the instrument response function (which can be non-linear), $N$ is the noise covariance, and $S$ is the signal covariance. Combining these terms, the inference task is given by determining the information Hamiltonian $H(\phi|d) = -\ln P(\phi|d)$, which is the negative logarithm of the posterior.

\par One of the main strengths of IFT is its rigorous handling of uncertainty. Rather than producing a single point estimate, the IFT framework can characterize the full posterior probability distribution, which encapsulates all statistical information about the signal. The challenge becomes to manage this in high-dimensional and non-linear scenarios where the posterior is non-Gaussian and direct computation is not possible. To address this, advanced computational methods were developed in the context of IFT like Metric Gaussian Variational Inference \citep[MGVI, ][]{knollmuller2020metricgaussianvariationalinference}. MGVI works by approximating the true posterior with a Gaussian distribution, which achieves a significant speedup and higher accuracy compared to traditional methods. This kind of scalability is essential for processing the large data volumes from next-generation observatories like SKAO, where computational performance is a key constraint.

\par To overcome the limitations of a simple Gaussian approximation, more advanced methods like Geometric Variational Inference (geoVI) were developed \citep{frank2021geometric}. The geoVI algorithm uses tools like normalizing flows to learn a non-linear transformation from a simple base distribution into a much more complex one. This allows the variational approximation to capture intricate features of the true posterior, such as non-Gaussian posteriors or non-linear degeneracies. Although geoVI is more computationally demanding than MGVI, it yields a more accurate and robust uncertainty characterization and allows for more complex probability distributions.

\par The applications of IFT are diverse, ranging from the reconstruction of cosmological large-scale structures \citep{jasche2010bayesian} to mapping the Galactic interstellar medium \citep{edenhofer2024astrometric}. In the field of astroparticle physics, it has been used to reconstruct radio pulses induced by cosmic rays and neutrinos \citep{welling2021cosmic, strahnz2024constraining}. Recent work has explored the use of IFT for air shower reconstruction, using techniques such as parameterized forward models for radio fluence-based reconstructions \citep{terveer2025reconstructing} and template synthesis for the full electric field \citep{watanabe2025novel}. These highlight IFT's potential for cosmic-ray physics with SKA-Low.

\par For this work, we aim to highlight current and future cosmic-ray reconstruction approaches using IFT. These approaches are currently tested on LOFAR and are being developed for use on SKA-Low data. Our implementation relies on the Python framework Numerical Information Field Theory (NIFTy,   \citealp{edenhofer2024niftyre, arras2019nifty, steininger2019nifty, selig2013nifty}). NIFTy simplifies the construction of IFT models and uses Google's JAX library \citep{bradbury2018jax} as its computational backend. This combination is particularly powerful, as JAX provides automatic differentiation and just-in-time (JIT) compilation, enabling execution on  GPUs—a critical requirement for processing data at the scale of the SKA-Low.

\section{Holistic reconstruction using IFT for LOFAR}
\label{sec:HolisticRecoLOFAR}

With prior reconstruction approaches fitting shower parameters separately as introduced in Section \ref{sec:CRDetection}, IFT allows us to reconstruct the shower as a whole based on a parametrized forward model. This means we can use all information to reconstruct all parameters simultaneously, while keeping the reconstruction runtime of $\mathcal{O}($\SI{1}{\hour}$)$ on 1 CPU. Ideally, we would reconstruct the whole electric field at all shower footprint positions, as this encodes all the shower parameters of interest. For now this method has been developed for the direct application in LOFAR and in the 30-\SI{80}{MHz} band. The strongly resonant antennas in LOFAR make the reconstruction of a frequency spectrum across its bandwidth challenging.
Instead, we reconstruct based on a combination of a fluence model and a wavefront model, to probe the feasibility of a holistic reconstruction with IFT, which could then be further expanded to SKA-Low, requiring a thorough understanding of the system response.  

\subsection{Fluence model}
The fluence is a measure of electromagnetic energy per unit area. At a given position $\vec{r}_\text{ant}$ in the shower plane, it is defined by Eq. \eqref{fluence}, defined within a signal window $T$, as shown below:
\begin{equation}\label{fluence}
F(\vec{r}_\mathrm{ant}) = \epsilon_0 c \sum_\mathrm{pol} \int_T \abs{E_\mathrm{pol} (t, \vec{r}_\text{ant})}^2\dd{t},
\end{equation}
where $E_\mathrm{pol}$ is the electric field trace in each polarization, which varies as a function of time and antenna position.

The radio fluence footprint of an air shower exhibits radial asymmetry due to the interference between the geomagnetic and charge excess effects, resulting in the prominent "bean-shape" of the footprint (see Fig. \ref{fig:fluence_reco_example} top center).
This behavior has been parameterized by the geoceLDF parameterization for near-vertical showers with $\theta<$ \SI{60}{\degree} \citep{GLASER201964}. The parameterization calculates both emissions separately for a set of cosmic ray parameters ($E_\text{rad}, \ D_{X_\text{max}}, \ \theta, \ \phi, \ X_\text{core}, \ Y_\text{core}$) while using information on the local magnetic field strength and atmospheric conditions and returns the radio footprint for the required observational height in shower coordinates ($\vec{e}_{v\times B}, \vec{e}_{v\times (v\times B)}$). 
We modified the implementation, making the code differentiable for the inference and adding convenient transformations: For one, a transformation between $D_{X_\text{max}}$ and $X_\text{max}$ via the relation 
\begin{equation}
    D_{X_{\text{max}}} = \frac{X_{\text{atm}}}{ \cos(\theta)} - X_{\text{max}}
\end{equation}
to make the model dependent on $X_\text{max}$ rather than $D_{X_{\text{max}}}$. As the reconstruction will optimize the angle of incidence simultaneously to the $X_\text{max}$ and all other parameters, and the transformation between $X_\text{max}$ and $D_{X_\text{max}}$ depends on the zenith angle, including it in the forward model provides more information for the inference.

Additionally, we added a transformation between the shower coordinates ($\vec{e}_{v\times B}, \vec{e}_{v\times (v\times B)}$) and the ground coordinates ($X,Y$). This has two advantages: It means the forward model outputs ground coordinates, so we can always use the antenna ground coordinates as reconstruction input, but it also means the forward model contains even more information on the angle of incidence.

\subsection{Wavefront model} \label{sec:wavefront}

The arrival time of the radio signal at the antennas depends on the angle of incidence of the shower, however the wavefront is also subtly shaped by the shower maximum $X_\text{max}$ \citep{Apel_2014}. We use the parameterization that was based on CoREAS simulations for our forward model.
We define our arrival time at an antenna $i$ of distance $d_i = \sqrt{(x'_i)^2 + (y'_i)^2}$ from the shower core based on a global time offset $t_0$, which defines the arrival time at the shower core, and a geometric time delay $\tau_{\text{geo}, i}$:
\begin{equation}
t_{\text{arrival}, i} = t_0 + \tau_{\text{geo}, i}.
\end{equation}
The time delay is then calculated, as in \cite{Apel_2014}, as
\begin{equation}
\tau_{\text{geo}, i} = \frac{1}{c} \left( \sqrt{(d_i \sin\rho)^2 + (c \cdot b_{\text{offset}})^2} + z'_{s,i} \cos\rho\right)
\end{equation}
where
\begin{equation}
\rho = \frac{X_{\text{max}}}{C_{X_{\text{max}}, \rho}} \cdot \cos^{\gamma_{\text{zen}}}(\theta_{\text{zen}}).
\end{equation}
As in the reference, $b_{\text{offset}}$ is fixed at \SI{-3}{ns}. For $\gamma_{\text{zen}}$, $C_{X_{\text{max}}, \rho}$ we re-fitted the values on LOFAR simulations and found a Gaussian prior of $\gamma_{\text{zen}}=\mathcal{G}(1.465, 0.292)$, and for  $C_{X_{\text{max}}, \rho}$ in units of \si{\gram\per\centi\meter\squared} we found a dependence on the slant distance to shower maximum $D_{X_\text{max}}$, to which we fit a third-degree polynomial as represented in Tab \ref{tab:poly_fit}.

\begin{table}
    \centering
    \renewcommand{\arraystretch}{1.3}
    \begin{tabular}{c|l}
         \textbf{Polynomial Term} & \textbf{Coefficient Value} \\ \hline
         $p_3 \cdot x^3$ & \SI{-8.955e-5}{(\gram\per\centi\meter\squared)^{-2}} \\
         $p_2 \cdot x^2$ & \SI{ 1.873e-1}{(\gram\per\centi\meter\squared)^{-1}} \\
         $p_1 \cdot x^1$ & \num{-1.302e2} \\
         $p_0$           & \SI{ 5.338e4}{\gram\per\centi\meter\squared}
    \end{tabular}
    \caption{Coefficients of the third-degree polynomial fit $C(x) = p_3 x^3 + p_2 x^2 + p_1 x + p_0$, where $x = D_{X_{\mathrm{max}}}$ in \unit{\gram\per\centi\meter\squared}.}
    \label{tab:poly_fit}
\end{table}

Like the fluence model, we implemented the wavefront model as a differentiable forward model which produces signal times at ground positions for a given parameter set ($t_0$, $X_\text{max}$, $\theta$, $\phi$, $X_\text{core}$, $Y_\text{core}$) and can thus also optimize for these parameters.

\subsection{Fluence-based reconstruction for LOFAR} \label{sec:fluence_only}
First, we tested only the fluence model. We used a set of pre-existing CoREAS simulations for these tests. The simulation set consists of 208 iron and 857 proton primaries with energies ranging from $10^{16}$--$10^{18.5}\,\si{eV}$, $X_\text{max}$ values from $550$--$1050\,\si{\gram\per\centi\meter\squared}$, azimuth angles within \ang{0} - \ang{360} and zenith angles within \ang{0} - \ang{45}.

Using the LOFAR implementation within the NuRadio-framework \citep{Glaser2019} and Fourier-based interpolation \citep{corstanje_high-precision_2023}, we generate electric field traces for all LOFAR antenna positions with a random core location uniformly distributed within [-100,100] m, from which we calculate the fluence per antenna. We then take measured fluence noise from existing LOFAR data and add this noise to the fluence values to produce realistic fluence data.

The ground positions of the antennas as well as the noisy fluence values serve as data input for the reconstruction. For the parameter priors we use information that, for real data, we would get from the particle detectors. Based on data of the particle detectors at LOFAR (LORA) alone we know the angle of incidence up to \SI{0.7}{\degree} and the core location up to \SI{6}{m}, given a shower with a core within \SI{150}{m} from the LORA center. We adopt a more conservative prior standard deviation of \SI{2}{\degree} for the arrival direction (zenith and azimuth) and \SI{15}{m} for the core location \citep{THOUDAM2014339}. Within this standard deviation we add a random fluctuation to the truth arrival direction and core value to simulate how real data might behave, and use the resulting value as mean value for our parameter prior distribution. Furthermore, we use uninformative, uniform priors for the depth of shower maximum and the radiation energy: $X_\text{max} \sim \mathcal{U}(550, 1050)\,\si{\gram\per\centi\meter\squared}$ and $E_\text{rad} \sim \mathcal{U}(10^4, 10^8)\,\si{\electronvolt}$.

As the mean of the fluence noise primarily stems from the Galaxy and is per definition always positive, we model the Galaxy as an additional log-normally distributed signal sharply peaked at the observed noise mean. Our noise model then remains a Gaussian distribution centered around 0 with the observed standard deviation of the LOFAR fluence noise. This approximation sufficiently describes the nature of the fluence noise.

We perform the inference using the geoVI algorithm \citep{frank2021geometric}.
We filter out 319 of the reconstructed events with a peak signal-to-noise ratio (ratio between the maximum fluence value across all antennas and the noise standard deviation, SNR) less than 3, as for these events a pulse would be almost impossible to find within the noise. Furthermore, we reject failed reconstructions with a strict $\chi^2 \in [0.85,1.05]$ cut. This reduces our successful reconstructions further to 352 showers. 

\begin{figure}[t]
    \centering
    \includegraphics[width=0.99\linewidth]{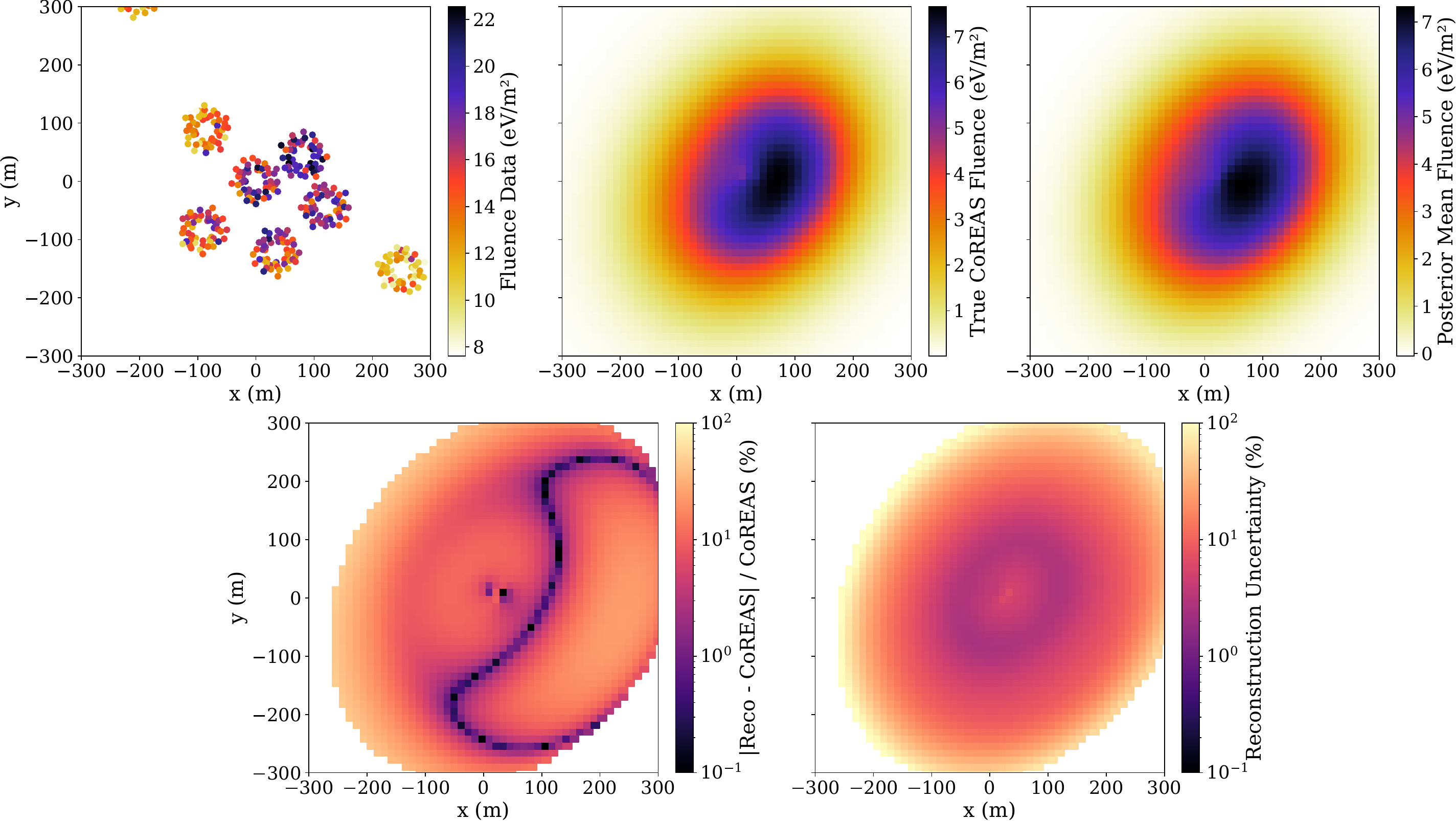}
    \caption{
    An example of the air shower reconstruction for a proton primary with an electromagnetic energy ($E_{\text{rad}}$) of \SI{6.61e5}{\electronvolt} and a shower maximum ($X_{\text{max}}$) at \SI{731.6}{\gram\per\centi\meter\squared}. The shower arrived from a zenith angle of \ang{37.0} and an azimuth of \ang{52.4}, with its core at $(\SI{21.5}{\meter}, \SI{7.7}{\meter})$. The event had a peak signal-to-noise ratio (SNR) of 3.5.
    The top row displays the input data and compares the true CoREAS radio footprint with the mean of the posterior reconstruction. 
    The bottom row shows the point-wise absolute difference and the relative uncertainty of the reconstruction, respectively. 
    The reconstructed parameters are $E_{\text{rad}} = \SI{5.50(42)e5}{\electronvolt}$ and $X_{\text{max}} = \SI{716(21)}{\gram\per\centi\meter\squared}$.
    }
    \label{fig:fluence_reco_example}
\end{figure}

An example of a successful reconstruction is shown in Fig. \ref{fig:fluence_reco_example}. It showcases both the input data as well as the true fluence from CoREAS and the posterior mean fluence in the top row. Additionally, the posterior standard deviation (reconstruction uncertainty) and the absolute difference between ground truth and reconstruction are shown in the bottom row.

\begin{figure}[t]
    \centering
	\includegraphics[width=0.9\columnwidth]{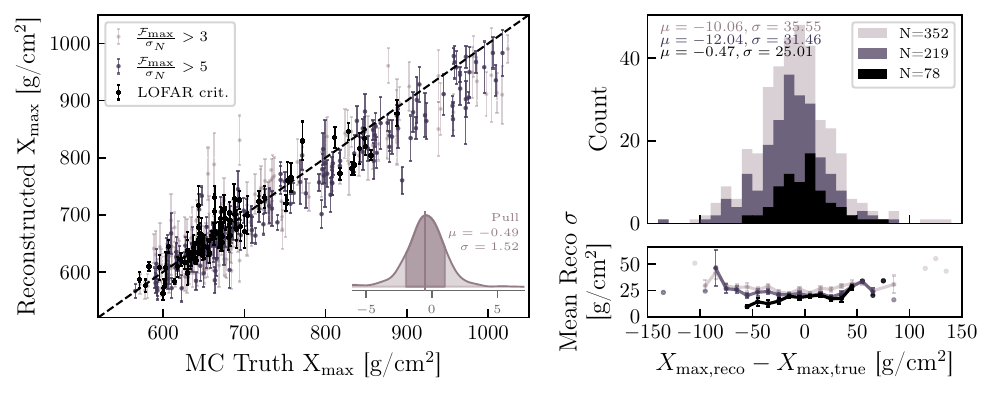}
    \includegraphics[width=0.9\columnwidth]{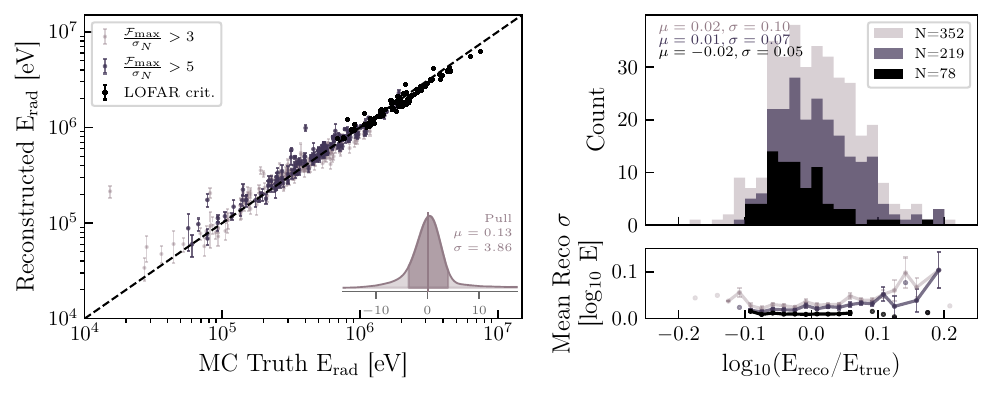}
    \caption{Fluence-based reconstruction performance for shower maximum (top row) and radiation energy (bottom row). For each parameter, the left plot shows the reconstructed vs.\ Monte Carlo true value, with an inset detailing the pull distribution. The right plot shows the resolution, with the distribution of reconstruction errors (top) and the mean reconstruction uncertainty per bin (bottom). The unconnected dots in the bottom right panels correspond to bins with only one value. Different colors represent different selection criteria as detailed in the legend: $\frac{\mathcal{F}_\mathrm{max}}{\sigma_N}$ refers to the ratio between the maximum fluence value across the data and the noise standard deviation. The LOFAR criterion refers to the selection criterion previously used in analyses, where at least one station with more than \SI{50}{\percent} of antennas are required to have a SNR$>6$.}
    \label{fig:fluenceOnly_xmax_energy_plot}
\end{figure}

Our results are split into three signal classes: Low SNR events with $3 < \text{SNR}_\text{peak} < 5$, high SNR events with $\text{SNR}_\text{peak} > 5$, and a third class of events where at least half of the antennas in each station had an $\text{SNR} > 6$ (LOFAR criterion). This third class is based on the quality cuts taken by previous LOFAR analyses as described in \cite{corstanje_depth_2021} and enables easier comparison of this method with the state-of-the-art. Figure \ref{fig:fluenceOnly_xmax_energy_plot} shows the results of the reconstruction. 

For lower SNR events the method shows a bias in $X_\text{max}$, however for the events passing the LOFAR criterion the bias disappears. For the LOFAR criterion dataset we calculate the Root Mean Square Error (RMSE) as measure of the overall performance of the reconstruction via $\mathrm{RMSE}=\sqrt{(\mathrm{bias})^2+(\mathrm{resolution})^2}$, which yields an $\text{RSME}_{\text{X}_\text{max}} = 25.01$ \si{\gram\per\centi\meter\squared}. The radiation energy reconstruction systematically underestimates the energy by 4.5\% on average, with a 1-sigma resolution of $^{+12.2}_{-10.9}\%$. The true radiation energy in the 30-\SI{80}{MHz} band can be difficult to determine due to artifacts produced during interpolation, however, which might be the reason for the apparent bias.

This first test of a radio footprint reconstruction showcases how IFT can be used as a fast alternative to conventional methods, with the added advantage of being able to reconstruct even lower SNR events and thus extending the energy range currently accessible by the standard reconstruction, albeit introducing a small bias in $X_\text{max}$.

\subsection{Combined Fluence-wavefront reconstruction}
A bias in $X_\text{max}$ can occur especially at low SNR, where noise dominates and the core position or arrival direction are harder to constrain for fluence data only. To counteract this, we introduce a second information source to our model, the signal time (see Sec.\,\ref{sec:wavefront}), which exhibits a second-order dependence on the shower maximum and core position and a first-order dependence on the signal direction.

The method for the combined reconstruction is described in detail in \citep{terveer2026}. We will summarize the important steps:

We build our combined signal model now by stacking fluence and timing models which now share the same prior distributions for shower parameters, yielding a model vector shape of $2N$ with $N$ the number of ground positions. Our data for reconstruction has the same shape, we calculate the fluence values and the timing values in each antenna and stack them before reconstruction. We use a set of CoREAS simulations and interpolate them to LOFAR antenna positions as before, this time using a less constraining condition for the core position, drawing it from $\mathcal{N}(0,100~\mathrm{m})$. Next, we apply the LOFAR antenna response to the electric field traces to obtain voltage traces for each antenna. From a library of measured pure noise at LOFAR, we take noise voltage traces and add them to the voltage signal traces to best simulate how real data will look like. For all following data processing steps we randomize the core position and arrival direction to simulate that the exact value will not be known for real data.

The additional challenge as compared to the fluence-based reconstruction in Sec.\,\ref{sec:fluence_only} is pulse finding. Before, we used the true fluence and added measured fluence noise, which meant that even for the low SNR events, the true fluence was always completely included in the noisy data. Additionally, the fluence-only reconstruction is less sensitive to whether the correct pulse time is found in the low signal antennas. In reality however, finding the pulse time for low SNR events is challenging. For that reason, we first identify a region of interest per antenna (ROI) by using an initial direction guess (randomized from the true direction) and performing a station-wide beamforming for $\pm$ \SI{10}{\degree} around that guess. With the ROI defined, we then search for a maximum of the Hilbert envelope of the trace. This is used as the signal time.  

The signal times that result from this hybrid approach are then cleaned in the next step. This is important, as for low SNR antennas the signal time might stem from a noise peak and be hundreds of nanoseconds off from the true time of arrival. Thus, we perform a local outlier rejection. Each antenna is compared to its 10 nearest neighbors by calculating the median and the standard deviation of all their timing values. If the antenna timing deviates by more than $2\sigma$ from the median value, the timing value is removed for the reconstruction. This pre-processing of the timing data is necessary as any strong deviations from the true timing values will massively impact the reconstruction performance.

The fluence values are calculated in a window of \SI{300}{ns} around the identified pulse times, by applying the inverse antenna response matrix (Eq. \eqref{eq:antenna_response}) to obtain electric fields and integrating over the signal window via Eq. \eqref{fluence}.
For the fluence-based reconstruction described above, including points with no signal was important for the reconstruction performance. However, since fluence and timing data need to have the same shape, this is not possible if many low-signal timing points are removed. For this reason, for the six most central stations (the LOFAR "superterp" stations), "bad" timing points are not removed but rather demoted. This means their value is kept but their noise covariance for the reconstruction is effectively set to zero, removing their impact on the reconstruction. For these points, the fluence is calculated from a time interval estimated by a plane-wave fit.

Estimating the noise covariance for the fluence data is simply done by taking off-signal intervals from the event and calculating the mean and standard deviation of the fluence noise. Additionally, we add a \SI{10}{\percent} systematic noise term to all fluence points, stemming from the discrete sampling and an estimation of detector calibration systematics.

The timing noise is more complicated, as it cannot directly be calculated from data and depends on the signal strength of the event. For high SNR events, the time of the maximum signal will usually be within $5-10$~\si{ns} (sampling time of LOFAR) of the true signal time, but for lower SNR events, due to the noise, the maximum might jump back or forth a few samples. Using the same timing noise estimate for all points is not an option, so we need to estimate the timing noise on a per-antenna basis using local clusters. Estimating the jitter via a global plane-wave fit is insufficient because its residuals, incorrectly combine systematic error from the wavefront curvature with the true random noise. We perform a polynomial fit per antenna that includes a radial curvature term, capturing both the plane-wave delay and the wavefront curvature. For each antenna $i$, we define a local cluster $C_i$ consisting of its $k=20$ nearest neighbors. Positions within this cluster are centered and normalized for numerical stability:
\begin{equation}
    \tilde{x}_j = \frac{x_j - \bar{x}_{C_i}}{\sigma_r}, \quad \tilde{y}_j = \frac{y_j - \bar{y}_{C_i}}{\sigma_r}, \quad \tilde{r}_j^2 = \tilde{x}_j^2 + \tilde{y}_j^2,
\end{equation}
where $\sigma_r$ is the standard deviation of radial distances within the cluster. A least-squares fit is performed using the design matrix $\mathbf{A} = [1, \tilde{x}, \tilde{y}, \tilde{r}^2]$, solving for coefficients $\vec{c}$ such that $t_j \approx \mathbf{A}_j \cdot \vec{c}$. The residuals $\epsilon_j = t_j - \mathbf{A}_j \cdot \vec{c}$ represent pure measurement noise, free from systematic curvature contributions. The per-antenna timing uncertainty is then:
\begin{equation}
    \sigma_{\tau,i} = \max\left( \text{std}(\{\epsilon_j\}_{j \in C_i}), \sigma_{\tau}^{\min} \right),
\end{equation}
where $\sigma_{\tau}^{\min} = \SI{1.5}{ns}$ is a floor value to prevent unrealistically small uncertainties. If a cluster has fewer than 8 antennas, a fallback uncertainty of \SI{5}{ns} is used.

\begin{figure}[t]
    \centering
	\includegraphics[width=0.9\columnwidth]{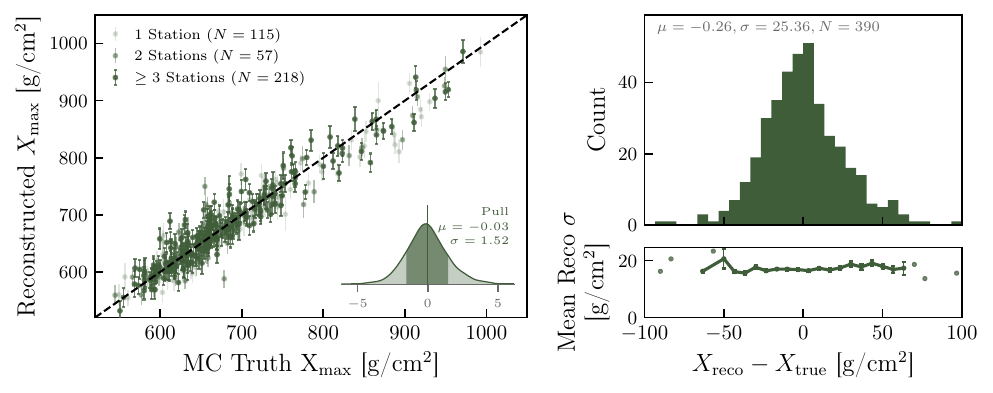}
    \includegraphics[width=0.9\columnwidth]{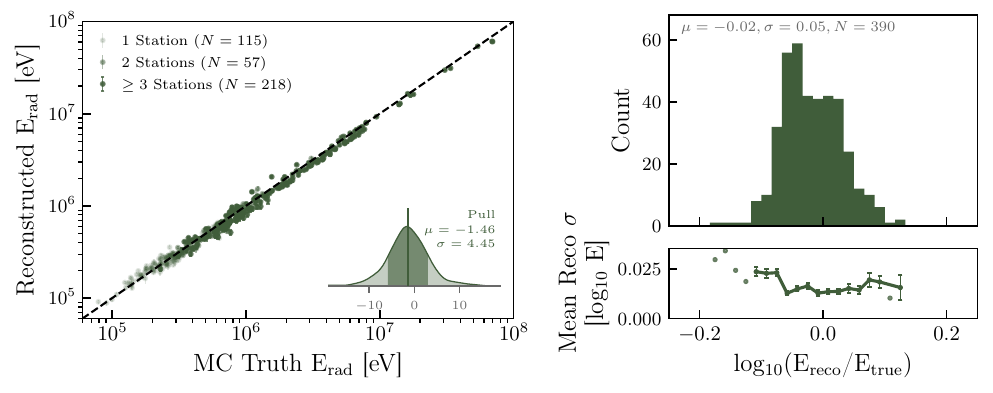}
    \caption{Performance of the radiation energy $E_\text{rad}$ and $X\text{max}$ reconstruction on CoREAS simulations. The left side displays both the reconstructed values compared to the CoREAS truth for all 390 successful reconstructions, as well as the pull from the reconstruction. The right side of the plots show histograms of the reconstruction errors at the top and how the reconstruction uncertainties correlate with reconstruction error. The different shades of green represent how many stations would have been "triggered" in this simulated event.}
    \label{fig:combined_xmax_energy_plot}
\end{figure}

For direction and core prior distributions we use the same approach of taking a random fluctuation around the LORA precision as prior mean and \ang{2} or \SI{20}{m} as standard deviation.

As for the fluence-based reconstruction we use the geoVI algorithm to perform the inference. Here, we used a set of 500 simulations with proton and iron primaries. An event was only reconstructed if at least in one station more than \SI{50}{\percent} of antennas had a fluence SNR $>6$. Post-reconstruction we keep only events with a reduced $\chi^2<1.25$ and an uncertainty of the reconstructed core smaller than \SI{10}{m}. With these cuts, around \SI{80}{\percent} of simulations remain in the final set. The performance is shown in Fig.~\ref{fig:combined_xmax_energy_plot}.

The plot shows a significant increase of $X_\text{max}$ and $E_\text{rad}$ reconstruction precision. We also find that the precision in zenith angle improves by a factor of 3 as compared to the fluence-only reconstruction. The bias in $X_\text{max}$ is reduced to less than 0.3 \grammage,, with an RMSE of 25.4 \grammage. 

The reconstruction systematically underestimates the energy by 4.5\% on average, with a 1-sigma resolution of $11.5\%$. We note that the Pull displayed in the plot indicates a systematic underestimation of uncertainties. This pull value can, for real data, be used to scale the methods uncertainties.

We already applied this method to a set of measured LOFAR events, with the IFT approach enabling the first simultaneous fitting of timing and fluence data \citep{terveer2026}, showcasing its potential on real data. 

Overall this work shows that combining different information from the data yields an increased reconstruction performance. We are working on adding information from the polarization of electric field and also particle detector data to this holistic reconstruction and expect the performance to improve even further. This work currently developed for LOFAR, will provide an excellent foundation for the usage in SKA-Low, as discussed in Section \ref{sec:Outlook}.

\section{Air Shower Profile Reconstruction with IFT}
\label{sec:ProfileRecoTS}

In Section~\ref{sec:HolisticRecoLOFAR}, we showed that IFT is fully capable of reconstructing relevant parameters for cosmic ray air showers within the context of the dense antenna layout of LOFAR, even when utilizing only the total energy fluence and timing at each antenna position. In this section, we highlight that we can go one step further and reconstruct the entire longitudinal evolution of the shower, which can provide us with more information the mass composition of the primary cosmic ray and reveal the interactions that occur within particles in the air shower (cf.~\citealp{Buitink01.2026.SKA}). This can be achieved by using \texttt{SMIET} \citep{desmet_smiet_2025}, which synthesizes individual electric field traces from the radio emission of any shower at each antenna location. After convolving the traces with the instrument response following the SKALA4.1 antenna model \citep{Bolli2020}, we use IFT to simultaneously utilize the available information from the pulse shape, amplitude, and relative timing between antennas in each voltage trace for reconstruction of the longitudinal profile of the shower. We showcase our reconstruction framework by benchmarking with results obtained from CoREAS simulations. While we apply our framework with an idealized star-shape antenna layout here, this approach can be expanded with the SKA-Low AA$^*$ antenna layout.

\subsection{Signal Model}

The signal model, in this context, encompasses all information from the shower observables that we infer to the electric fields observed from our model for the radio emission of the air shower. In the following subsections, we describe the implementation of each step of the signal model.

\subsubsection{Longitudinal Profile}

\begin{figure}
    \centering
    \includegraphics[width=0.7\linewidth]{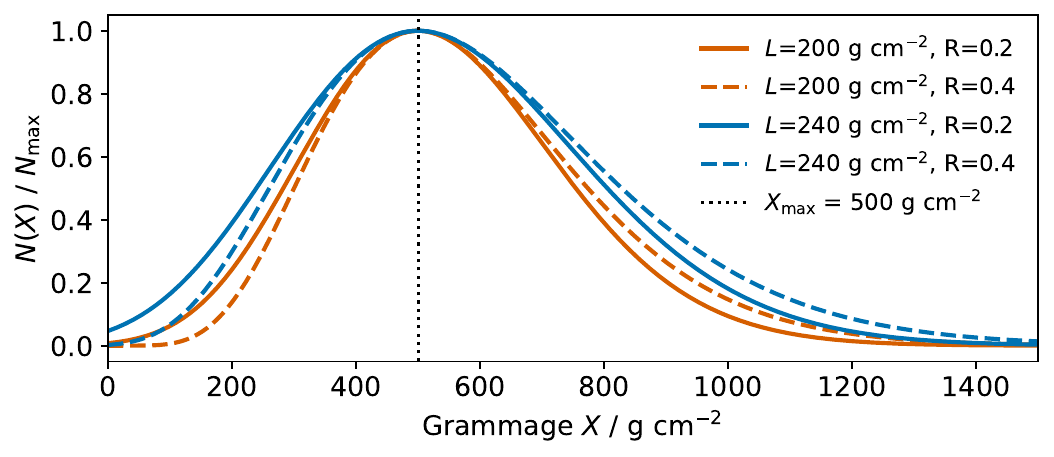}
    \caption{The longitudinal shower development parameterized using the Gaisser-Hillas function in the L-R-formalism. We take $X_\mathrm{max} = \SI{500}{\gram\per\centi\meter\squared}$ and normalise the function by $N_\mathrm{max}$.}
    \label{fig:gaisser_hillas_lr}
\end{figure}

The longitudinal profile describes the evolution of the air shower as it traverses through the atmosphere. The profile is typically defined in terms of atmospheric depth or ``grammage'' $X$ (in units of g cm$^{-2}$), describing the amount of matter traversed within the atmosphere. While several parameterizations of the longitudinal profile exist \citep{Griesen:1956, Gaisser:1977}, we opt to use the Gaisser-Hillas function in the L-R-formalism \citep{Andringa:2011zz} to describe the longitudinal profile
\begin{equation}
    N(X) = N_\mathrm{max} \exp\left(-\dfrac{X - X_\mathrm{max}}{RL}\right) \: \left(1 + \dfrac{R}{L} (X - X_\mathrm{max})\right)^{R^{-2}}.
    \label{eq:gaisser_hillas}
\end{equation}
where $N_\mathrm{max}$ is the number of electrons and positrons at the shower maximum and $X_\mathrm{max}$ is the atmospheric depth at $N_\mathrm{max}$. $L$ and $R$ are related to the width and skewness of the shower, respectively. Fig.~\ref{fig:gaisser_hillas_lr} shows the Gaisser-Hillas function, showing that larger values of $L$ increase the width of the shower, while increasing values of $R$ skew the shower towards larger depths. 

The shower parameters describing the longitudinal profile are sampled from a prior distribution, which encodes the current knowledge of each parameter. In practice, it is ideal to use weakly informative priors to describe the posterior without explicitly including information. However, as a first step, we use relationships known from current studies of air shower physics and encode these into the prior. Previous studies have shown that $X_\mathrm{max}$, $L$ and $R$ are strongly correlated with each other. As such, we model the combined distribution between these three parameters using a multivariate normal distribution, where the covariance matrix allows us to encode their correlations. \par 

\begin{figure}
    \centering
    \includegraphics[width=0.6\linewidth]{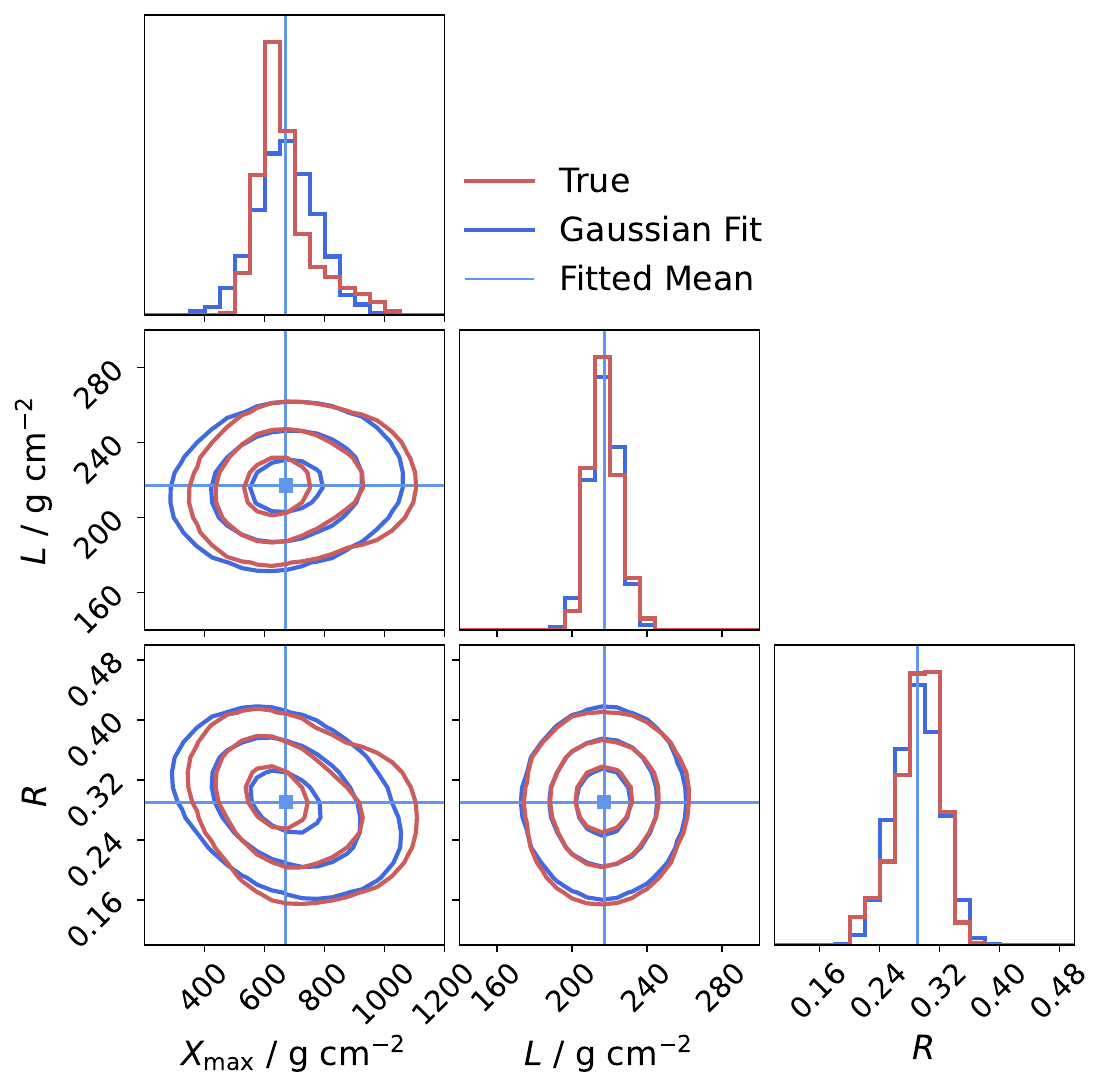}
    \caption{Corner plot of the truncated multivariate normal distribution used as our prior distribution (blue). We also overlay the true distribution obtained from CORSIKA simulations (red). The mean value from the fit are also shown. The contours represent the 1, 2, and 3$\sigma$ confidence level for a 2-D Gaussian distribution. Figure generated using \texttt{corner.py} \citep{corner}.}
    \label{fig:corner_plot_ghparams}
\end{figure}

To determine the mean and covariance matrix values, we take the longitudinal profile of $\sim$10,000 CORSIKA simulations \citep{Heck:1998vt}, generated with the hadronic interaction model QGSJETII-04 \citep{Ostapchenko2013} with environmental conditions adequate for the site of LOFAR. We fit each profile with Eq.~\ref{eq:gaisser_hillas} and perform another fit of the obtained parameters to a multivariate Gaussian to get an estimate of the mean and covariance matrix. In doing so, we apply physical limits to the shower parameters: 
\begin{equation*}
    X_\mathrm{max} \in [400, 1100], \quad L \in [200, 240], \quad R \in [0.2, 0.4],
\end{equation*}
so that our prior distribution is described through a truncated multivariate normal distribution. Fig.~\ref{fig:corner_plot_ghparams} shows the prior distribution for $X_\mathrm{max}, L$, and $R$ used in this work. As $N_\mathrm{max}$ is only weakly correlated with all other parameters that describe the longitudinal profile, we separately model with a lognormal distribution, i.e.~$\log_{10} N_\mathrm{max} \sim \mathrm{Normal}(8, 0.5)$, instead. Additionally, we use an Ornstein-Uhlenbeck process to capture numerical fluctuations that cannot be adequately captured from the parametric function \citep{Uhlenbeck:1930zz}. 

\subsubsection{Modeling Radio Emission from Air Showers}

To fully incorporate the physics describing the radio emission generated from air showers into the context of IFT, a fast, memory-efficient, and fully differentiable model is required. This can be achieved by using \texttt{SMIET}, a fully differential, semi-analytical model for radio emission of air showers \citep{desmet_smiet_2025, desmet_smiet_2025_zenodo}, which has been developed for this purpose. This model is based on the template synthesis approach \citep{desmet_proof_2024} where results from CoREAS simulations which rely only on microscopic features evaluated through first principles, are parameterized based on air-shower universality \citep{Lipari2009} to macroscopically describe the radio emission. In the following paragraphs, we outline the general idea of the model, and defer the readers to relevant publications \citep{desmet_proof_2024, desmet_smiet_2025} for more details.\par

The core idea of \texttt{SMIET} lies in the concept of shower universality, where, under re-parameterization of shower observables, the electromagnetic component of the air shower can be described similarly from each other \citep[e.g.][]{Lipari2009}. To utilize this concept, a set of parametric functions are generated through a large ensemble of CoREAS simulations, which are parameterized through the following observables:
\begin{itemize}
    \item the atmospheric slice relative to $X_\mathrm{max}$ : $\Delta X_\mathrm{max}^\mathrm{slice} = X - X_\mathrm{max}$, 
    \item the antenna viewing angle, i.e.\ the opening angle between the slice and each antenna : $\theta_\mathrm{ant}^v$, described in units of the local Cherenkov angle $\theta^C_\mathrm{slice}$, and
    \item the frequency at which the signal is observed : $f$,
\end{itemize}
for a given geometry $(\theta, \phi)$.
These functions are then further rescaled through environment- and slice-dependent parameters, such as:
\begin{itemize}
    \item the atmospheric density at each slice, $\rho_\mathrm{slice}$,
    \item the geomagnetic angle, $\alpha_\mathrm{GEO}$, i.e., the angle between the shower axis and the geomagnetic field,
    \item the refractive index at each slice, $n_\mathrm{slice}$ (inversely related to $\cos \theta_\mathrm{slice}^C$),
    \item the number of particles that emit radio waves in each slice, $N_\mathrm{slice}$, and
    \item the geometrical distance from each atmospheric slice to each antenna, $d_\mathrm{slice}$.
\end{itemize}
For convenience, we encode all such parameters as $p_\mathrm{slice} := (N_\mathrm{slice}, d_\mathrm{slice}, \theta_\mathrm{slice}^C, \rho_\mathrm{slice}, \alpha_\mathrm{GEO})$. The resulting amplitude spectrum for the geomagnetic (GEO) and charge-excess (CE) emission, $\tilde{A}_\mathrm{GEO/CE}$, is then parameterized for different values of $\Delta X_\mathrm{max}^\mathrm{slice}$ and $\theta_\mathrm{ant}^v$. 

We now use a single shower simulated from CoREAS at a given geometry $(\theta, \phi)$ that contains the emission spectrum at each atmospheric slice (which we call as the ``origin shower'') to generate a template to which we can then apply any longitudinal profile to get the radio emission at each slice. Using the slice- and environment-specific parameters from the origin shower, we rescale the emission spectrum from the origin shower with the parameterized spectrum to generate the template:
\begin{equation}
    A^\mathrm{template}_\mathrm{GEO/CE}(\theta_\mathrm{ant}^v, \Delta X_\mathrm{max}^\mathrm{slice}, f) = \left[\tilde{A}_\mathrm{GEO/CE}(\theta_\mathrm{ant}^v, \Delta X_\mathrm{max}^\mathrm{slice}, f; p_\mathrm{slice}^\mathrm{origin})\right]^{-1} \: * \: A^\mathrm{origin}_\mathrm{GEO/CE}(\theta_\mathrm{ant}^v, \Delta X_\mathrm{max}^\mathrm{slice}, f).
    \label{eq:template_spectra}
\end{equation}
In this way, the template not only describes the self-similar features through the parameterization but can also encapsulate microscopic features that cannot be parameterized within a single shower. \par 

With this template, we can now use any arbitrary longitudinal profile to synthesize the radio emission observed at each antenna. The emission spectrum of the shower (which we call the ``target shower'') is synthesized as follows:
\begin{equation}
    A^\mathrm{target}_\mathrm{GEO/CE}(\theta_\mathrm{ant}^v, \Delta X_\mathrm{max}^\mathrm{slice}, f) = \tilde{A}_\mathrm{GEO/CE}(\theta_\mathrm{ant}^v, \Delta X_\mathrm{max}^\mathrm{slice}, f; p_\mathrm{slice}^\mathrm{target}) \, * \, A^\mathrm{template}_\mathrm{GEO/CE}(\theta_\mathrm{ant}^v, \Delta X_\mathrm{max}^\mathrm{slice}, f)
    \label{eq:synthesised_spectra}
\end{equation}
where the longitudinal profile, $N(X)$, enters in $p_\mathrm{slice}^\mathrm{target}$ as the number of emitters in each slice. We note that it is also possible to parameterize the phase component of the emission appropriately within this framework, which also allows us to infer the arrival direction of the air shower (within a few degrees). However, we defer this implementation for future works, and focus on inference for a single geometry for now. Furthermore, the parameterized spectra are defined for fixed antenna positions that lie in an idealized star-shape antenna layout. As such, synthesis is only possible at the same fixed antenna positions. While there are techniques to interpolate the signal to arbitrary positions \citep{corstanje_high-precision_2023}, we opt to fix the antenna positions in this work and extend our model to synthesize signals at any position in future works. \par 

\begin{figure}
    \centering
    \includegraphics[width=0.69\linewidth]{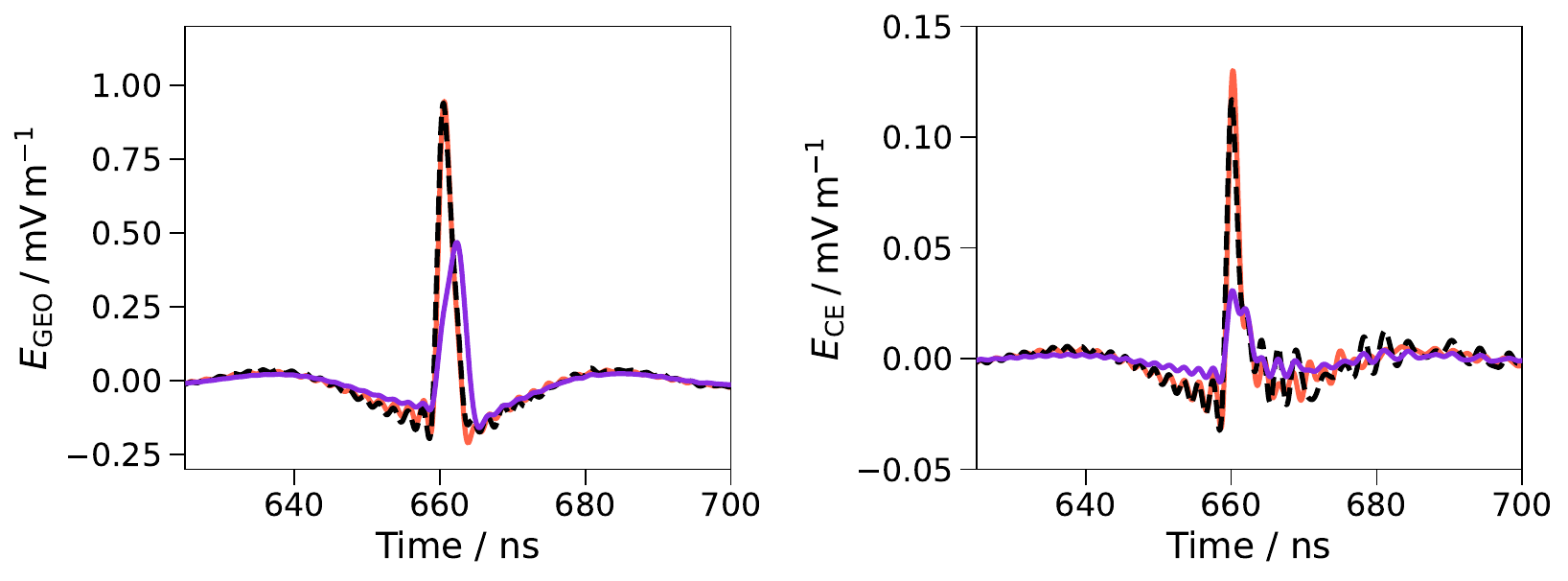}
    \includegraphics[width=0.3\linewidth]{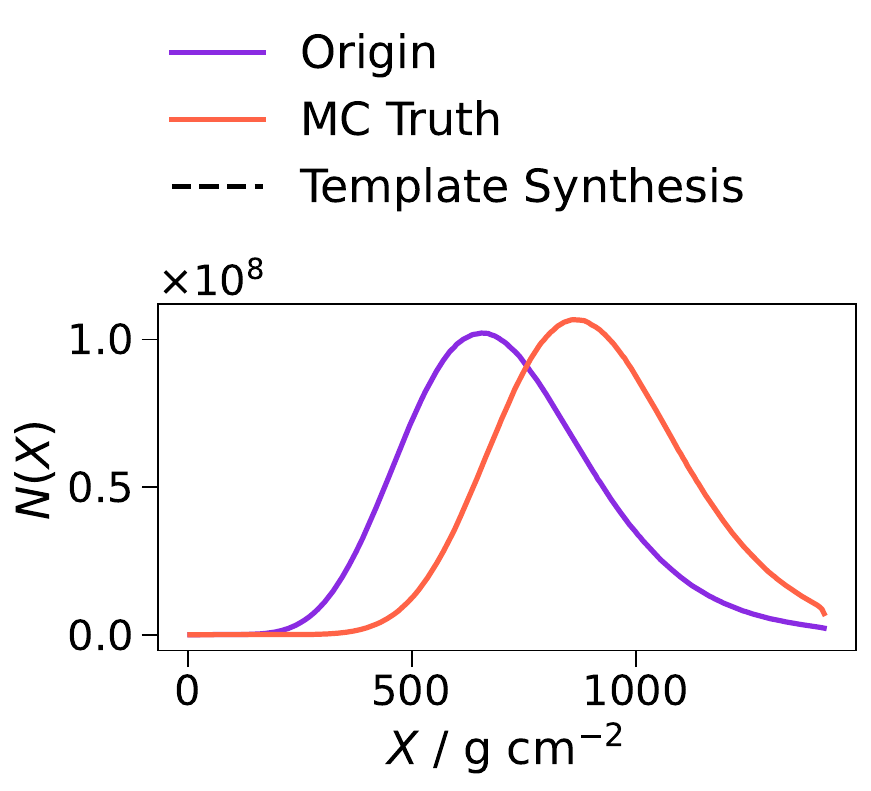}
    \caption{The electric field traces obtained from \texttt{SMIET} (dashed) and from CoREAS (orange) in the geomagnetic and charge-excess components. The traces are band pass filtered to 30 - 500 MHz. The traces were generated by using the same longitudinal profile (in orange). The longitudinal profile and electric field trace of the origin shower (in purple) used to generate the synthesised trace is also shown. }
    \label{fig:smiet_example_trace}
\end{figure}

Through \texttt{SMIET}, we are now able to generate the radio emission of cosmic ray air showers in a matter of seconds, orders of magnitude faster compared to microscopic Monte Carlo simulations. This approach also matches the accuracy of CoREAS within 6\% for showers synthesized with their maximum within 100 g cm$^{-2}$ from the $X_\mathrm{max}$ of the origin shower, and its performance has also been extensively tested under different atmospheric conditions, yielding similarly accurate results (see \citealp{desmet_smiet_2025}). To illustrate this, Fig.~\ref{fig:smiet_example_trace} shows an example trace of the electric field in the geomagnetic and charge-excess components generated with both CoREAS and \texttt{SMIET}, indicating excellent agreement with each other despite using an origin shower with a different longitudinal profile. 

In this work, we use \texttt{SMIET} to generate many instances of synthesized traces from each longitudinal profile, generated from the prior-sampled shower parameters. Each reconstruction utilizes a single origin shower with a fixed geometry and antenna positions. 

\subsection{Antenna Response and Noise Modelling}

The electric fields observed at each radio antenna are converted into a voltage signal, following the response function of the antenna. The response function, $\mathcal{H}$, depends on the observed frequency, electric field polarization, and the antenna orientation. The transformation from an observed electric field signal (in spherical coordinates) can be transformed into the voltage signal that each antenna observes using the following:
\begin{equation}
    \begin{bmatrix}
        V_X \\ V_Y 
    \end{bmatrix}
    = 
    \begin{bmatrix}
        \mathcal{H}_{X, \Theta} & \mathcal{H}_{X, \Phi} \\
        \mathcal{H}_{Y, \Theta} & \mathcal{H}_{Y, \Phi} \\
    \end{bmatrix}
    \cdot 
    \begin{bmatrix}
        E_\Theta \\ E_\Phi 
    \end{bmatrix}
\label{eq:antenna_response}
\end{equation}
where ($\Theta$, $\Phi$) are the polar and azimuthal components of the electric field, respectively\footnote{We use $\Theta$ and $\Phi$ as opposed to the standard notation $\theta$, $\phi$ to distinguish between the electric field components and the zenith and azimuthal angles of the air shower.}. Here, we use the SKALA4.1 antenna model \citep{Bolli2020}, which emulates the behavior of the SKALAs deployed at SKA-Low. These antennas are most sensitive in the 50-350 MHz region; most ideal for cosmic ray studies as their emission is strongest in the frequency bandwidth of 10 - 100 MHz. As such, each electric field signal generated from \texttt{SMIET} is transformed into a voltage trace using Eq.~\ref{eq:antenna_response}. \par 

 The likelihood defined in the IFT framework is defined through the noise, i.e.\ the difference between the observed data and the measured signal (as shown in Eq~\ref{eq:ift_likeli_prior}). As such, as opposed to standard algorithms, which focus on subtracting the sources of background noise that are measured in addition to the signal, noise sources must be included within the model. The most prominent source of background noise for radio emission of cosmic ray air showers is from the Galaxy, produces signals most dominantly from 30 - 100 MHz. Aside from this, locally produced instrumental noise from, e.g., TV towers, transmission cables, and other anthropogenic sources can affect the radio signal. Including all possible noise sources is highly non-trivial as we cannot easily identify all possible sources, and furthermore, these signals may or may not be differentiable, which is a necessary condition to be included in the model. As such, we model the noise by taking a Gaussian likelihood $\mathcal{N}(0, \sigma_N)$ with a noise root-mean-squared (RMS) amplitude (or equivalently the uncertainty) of $\sigma_N = \SI{7e-5}{\volt}$ in each time bin (i.e.\ each sample). This choice of the noise RMS is motivated to ensure that the signal-to-noise ratio (SNR), defined here to be the ratio between the maximum amplitude of the signal to the noise RMS, to be $> 1$. Through this simplified model, we attempt to encapsulate the possible sources of noise that can affect the signal. We plan to investigate a more sophisticated noise model, which also considers correlations of the noise sources between each time bin, in future works.

\subsection{Results}

To verify the performance and accuracy of our model, we benchmark our model by applying it to simulated data based on CoREAS. We take a subset of simulated datasets used in \cite{corstanje_depth_2021}, where approximately 20 CoREAS simulations per geometry, primary energy, and primary composition were performed, using environmental and atmospheric conditions from LOFAR. In this work, we choose seven event sets (i.e., geometry and primary energy), with proton as the primary composition for all simulations. This gives us a total of 142 simulations. Each simulated electric field trace is then convolved with the antenna response as in Eq.\ref{eq:antenna_response} to generate the voltage trace. We then include a randomized realization of noise following the noise model used within our framework to synthetically generate data, which is used to infer the relevant shower parameters by applying our model. \par 

\begin{figure}
    \centering
    \includegraphics[width=0.52\textwidth]{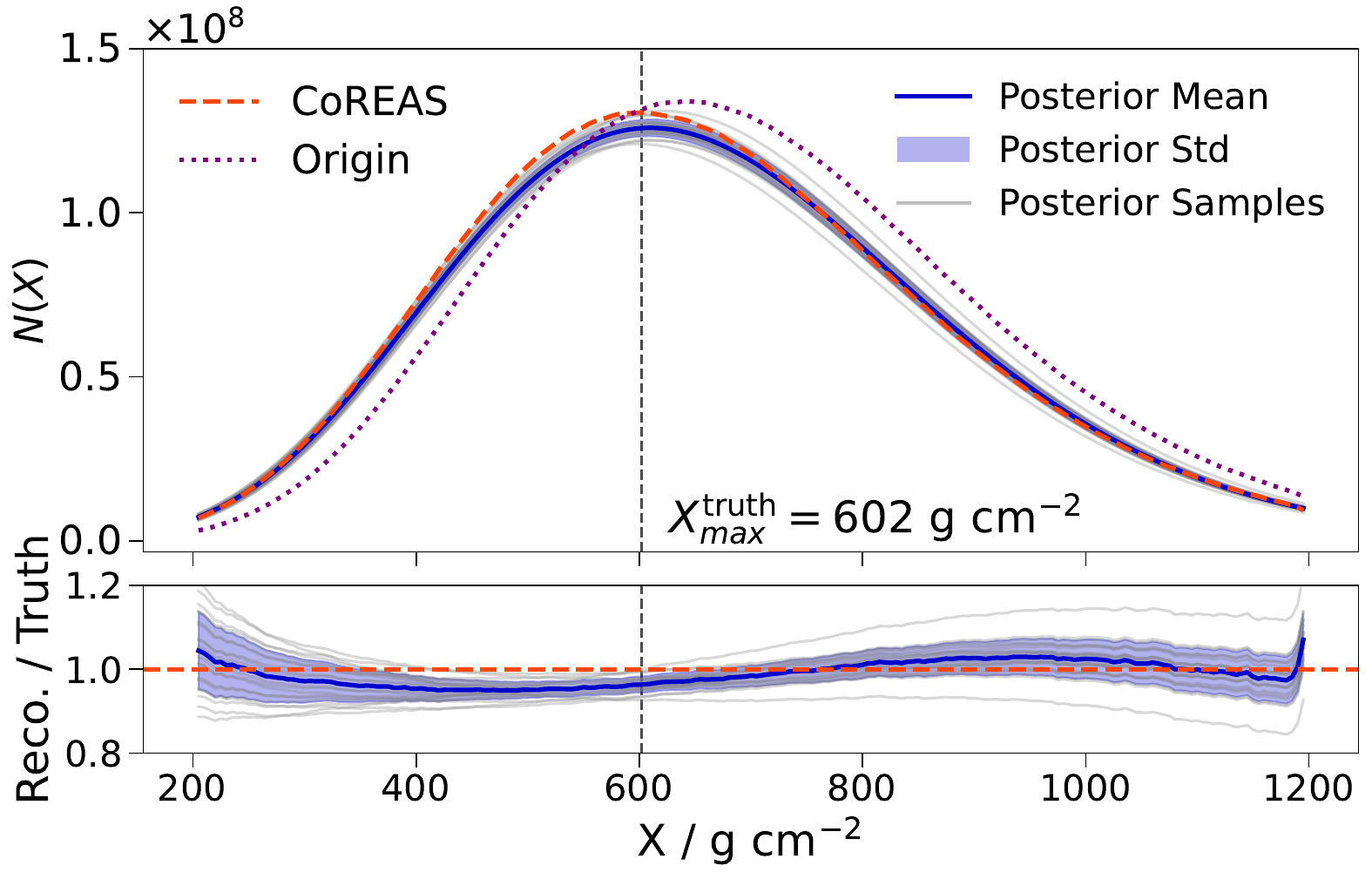}
    \includegraphics[width=0.43\textwidth]{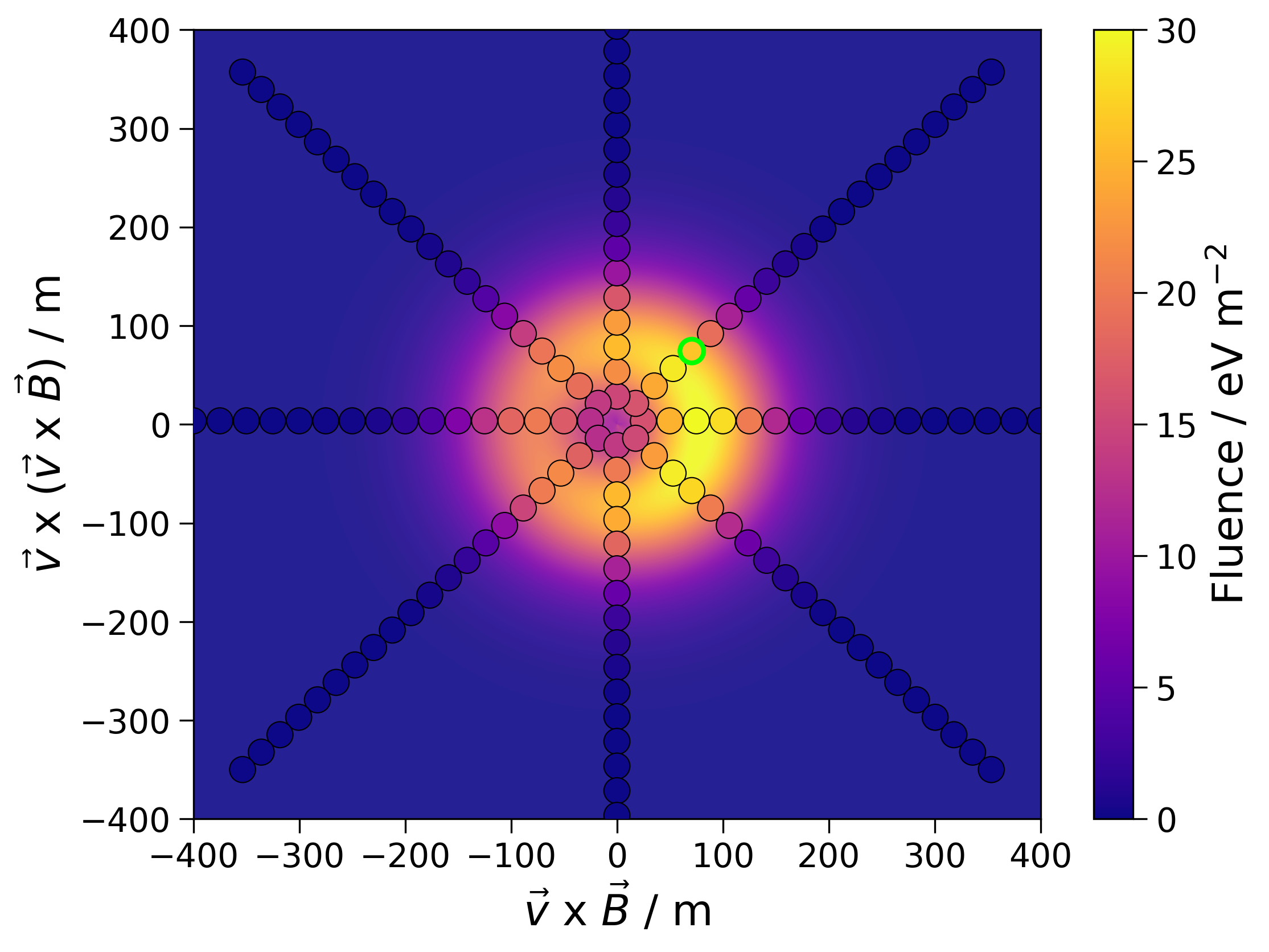} \\
    
    \includegraphics[width=0.49\textwidth]{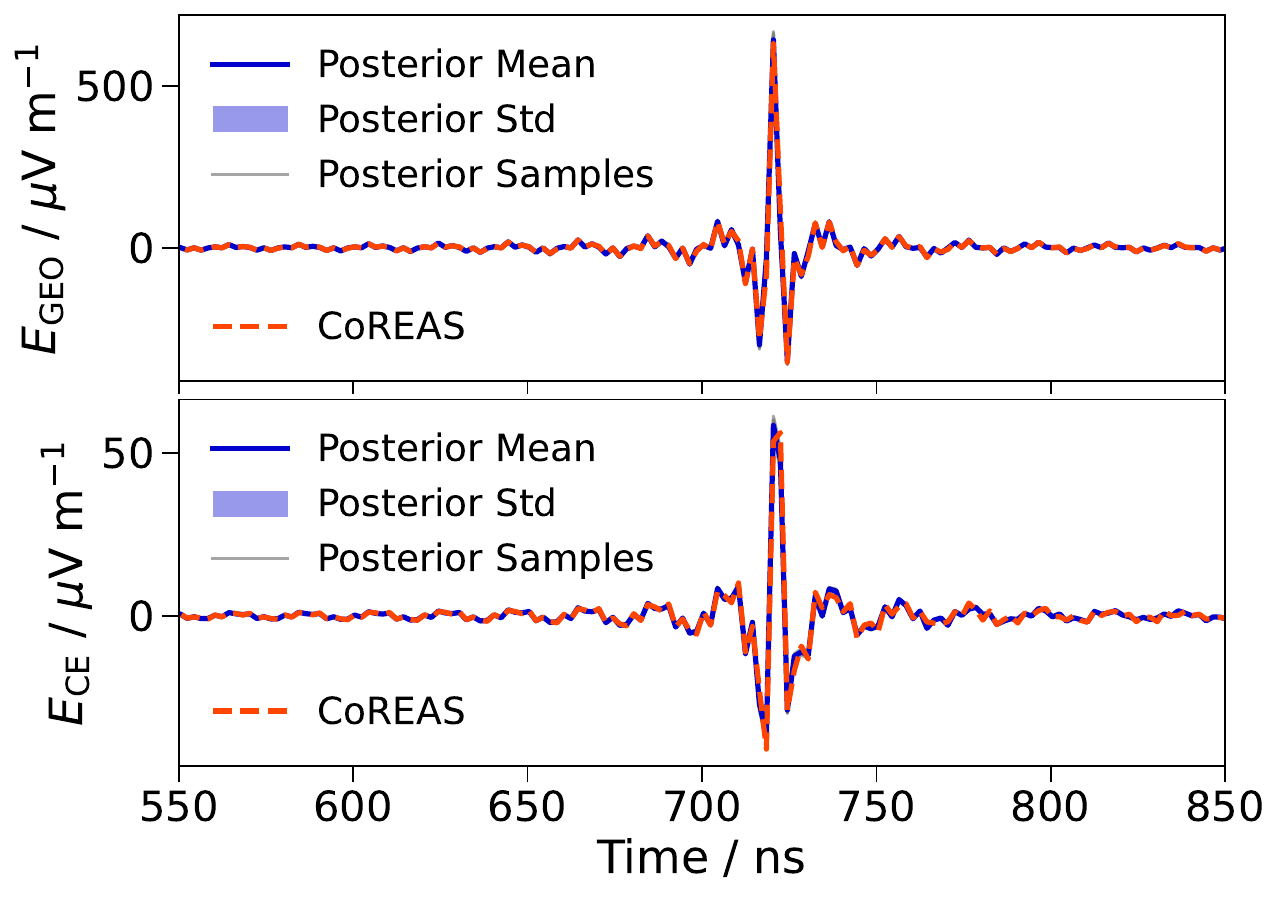}
    \includegraphics[width=0.49\textwidth]{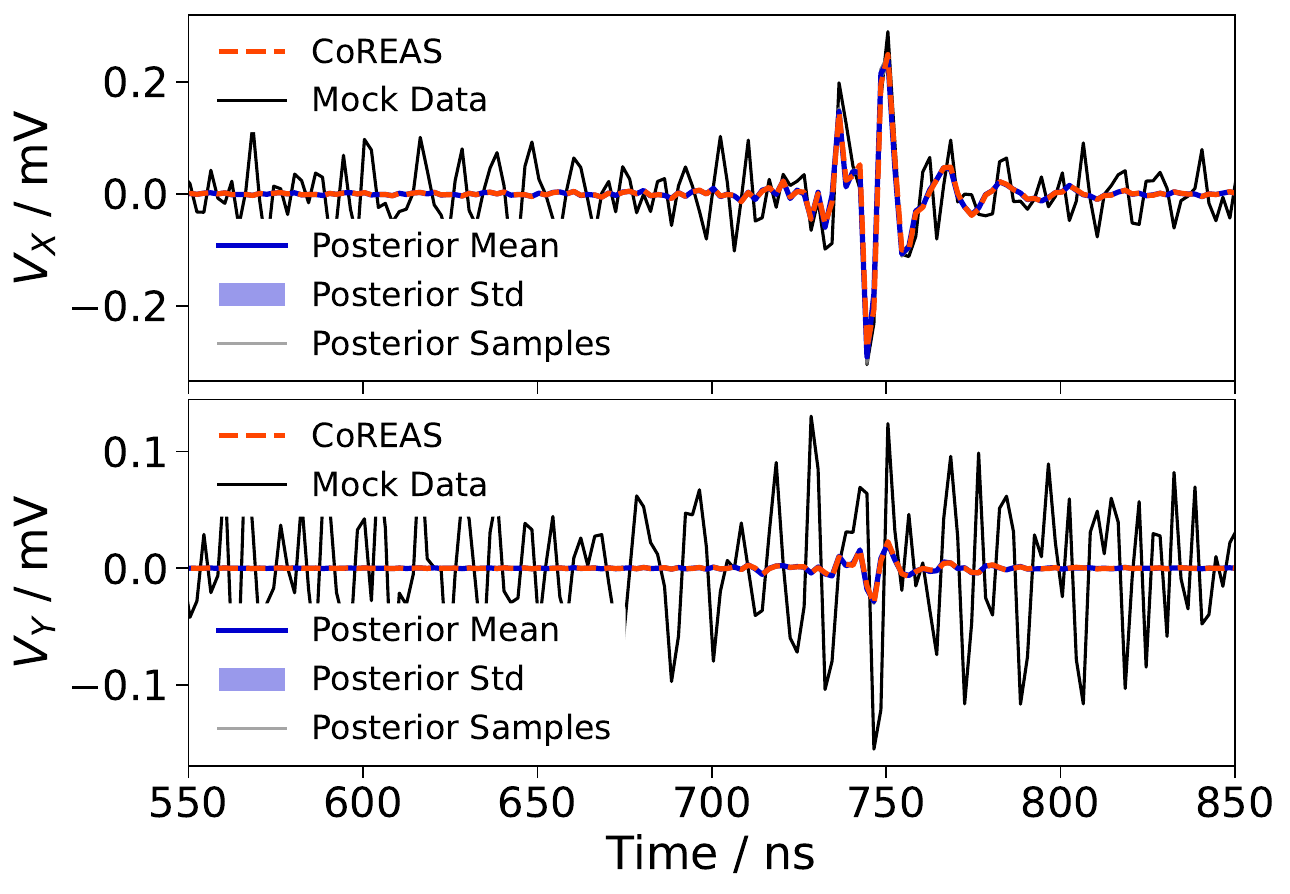}
    \includegraphics[width=0.48\textwidth]{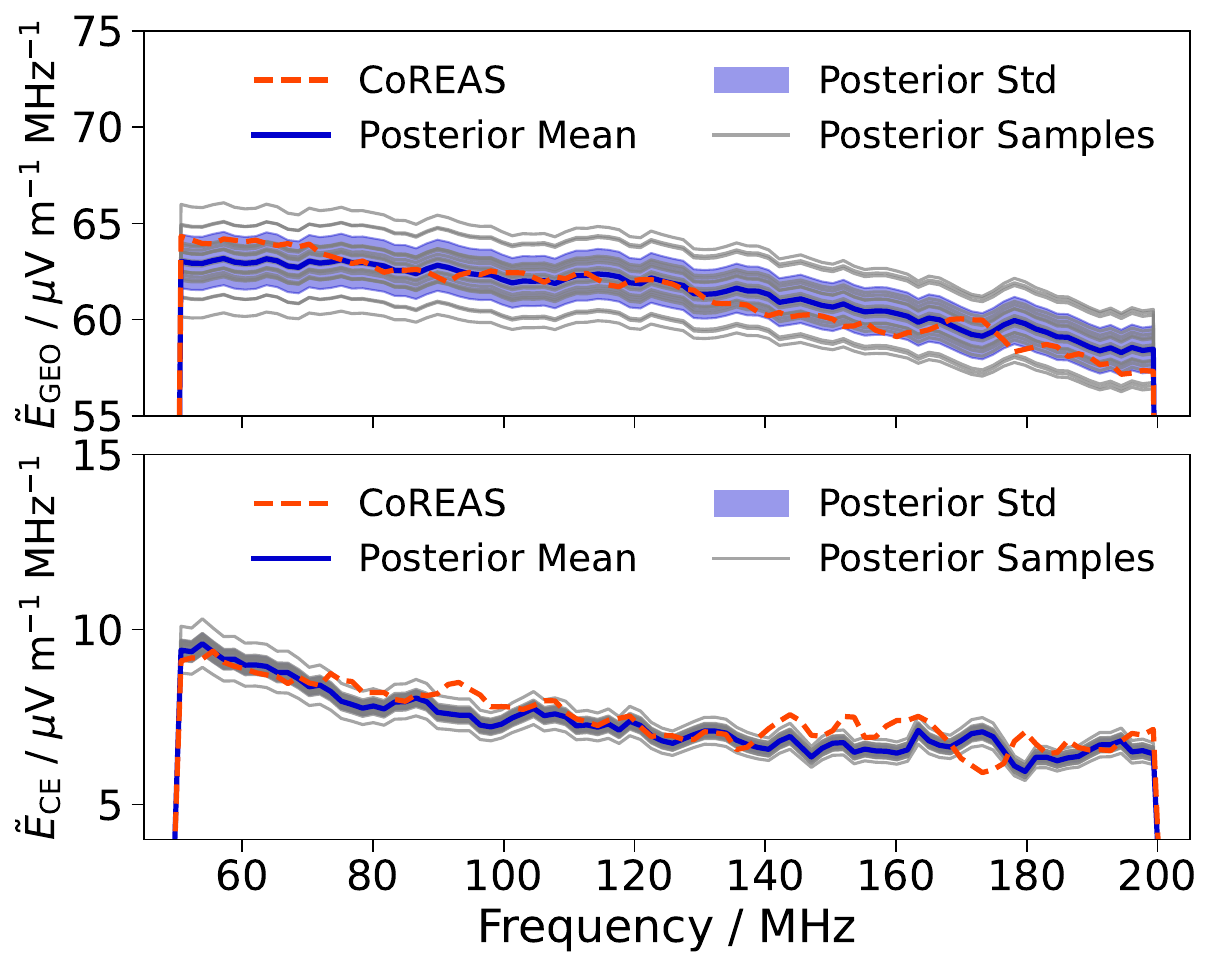} \quad
    \includegraphics[width=0.48\textwidth]{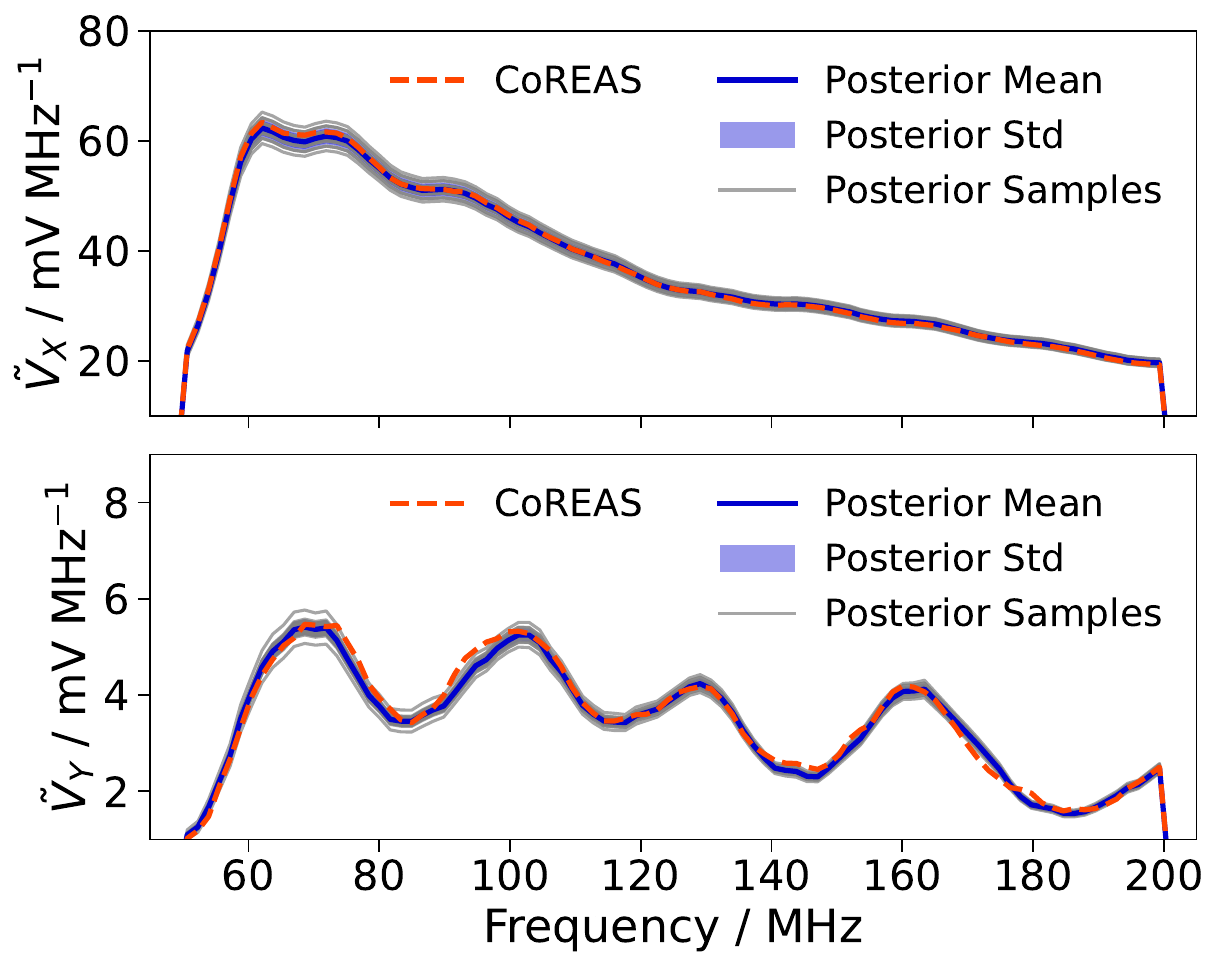}
    \caption{An example of an event reconstruction from this model with $E = \SI{2e17}{\electronvolt}$, $\theta = 28^\circ$, $\phi = 84^\circ$, and $X_\mathrm{max}^\mathrm{truth} = \SI{602}{\gram\per\centi\meter\squared}$. For each plot, we show the posterior mean and standard deviation (blue), true values from the CoREAS simulation (red, dashed), and each posterior sample used in the inference (gray). Top Left: The reconstructed longitudinal profile. The profile from the origin shower used in the template (purple, dotted) is also shown. The ratio between the reconstructed profile with those from CoREAS are shown in the bottom subplot.
    Top Right: The posterior mean of the energy fluence at each antenna (depicted by circles) in the shower plane. The interpolated true energy fluence is overlaid to show the qualitative agreement with the reconstructed results. The antenna position shown for the electric and voltage traces shown in the middle and bottom plots are indicated by the green circle.
    Middle Left: The reconstructed electric field traces in the geomagnetic and charge-excess polarizations (top and bottom respectively) at an antenna $d_\mathrm{core} = \SI{106}{\meter}$ away from the shower core. 
    Middle Right: The reconstructed voltage traces at the same antenna position in the X/Y polarization of the SKALA antenna. The mock data used for reconstruction is also shown. Bottom Left: The amplitude spectrum in frequency space for the electric field shown above. Bottom Right: Same as the bottom left, but for the voltage traces instead. }
    \label{fig:profile_event_reco}
\end{figure}

For each event set, we also generate a single origin shower, ensuring that $X_\mathrm{max}^\mathrm{origin} \geq \SI{640}{\gram\per\centi\meter\squared}$, which is used for \texttt{SMIET} within the forward model. We also perform our reconstruction within 50 - 200 MHz with a sampling rate of 500 MHz as the computational cost increases significantly when increasing the sampling frequency. For each reconstruction, we use the MGVI algorithm of \texttt{NIFTy} with 10 samples, and run each reconstruction for 10 iterations, such that we ensure that the fit has converged. \par 

Fig.~\ref{fig:profile_event_reco} shows an example of a reconstructed event with our framework. We highlight that, firstly, we are able to reconstruct the full longitudinal profile that, due to the posterior samples used to reconstruct the profile, yields a mean and standard deviation value (corresponding to a 1 $\sigma$ value for a Gaussian distribution) at each atmospheric bin. Comparing our results with the longitudinal profile from CoREAS, we observe that we are able to accurately reconstruct the profile at all bins in the atmosphere. The subsequent figures also show that we can reconstruct all other properties within the forward model (electric field and voltage traces) which show good qualitative agreement between the CoREAS simulated results at each level of our model. However, we observe from the amplitude spectrum in the $\vec{v}\times (\vec{v}\times\vec{B})$ component that minor fluctuations in the electric field signal due to shower-to-shower fluctuations or numerical noise, which cannot be directly observed in the voltage spectrum, are not adequately captured. Possible improvements, such as including fluctuation parameters within the electric field signal, can be applied to correct for this. \par 

\begin{figure}
    \centering
    \includegraphics[width=0.7\textwidth]{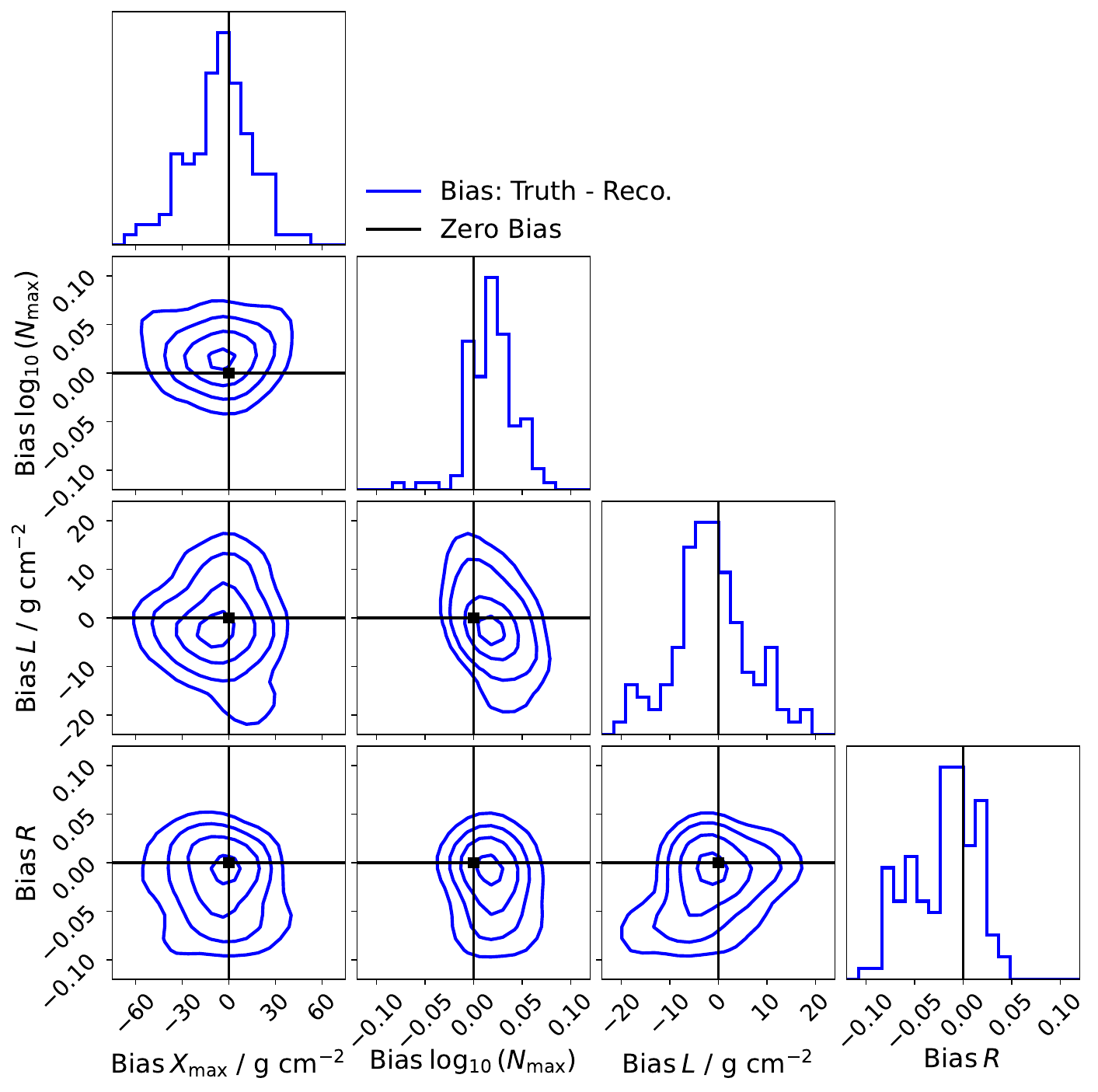}
    \caption{Distribution of the bias of the shower parameters $X_\mathrm{max}$, $\log_{10} (N_\mathrm{max})$, $L$ and $R$ for all events considered in this work. The contours show the 0.5, 1, 1.5, and 2$\sigma$ values following a 2-D Gaussian distribution. The point where the bias is zero is indicated by the black lines and dots. Figure generated using \texttt{corner.py}~\citep{corner}.}
    \label{fig:bias_gh_params}
\end{figure}

To verify our framework over all event sets, we show the distribution of the bias for each shower parameter in Fig.~\ref{fig:bias_gh_params}, taking the difference between the true values from CoREAS with the posterior mean from each reconstruction. We show that, in general, the reconstructed values are unbiased, albeit with a large variation. The precision of $X_\mathrm{max}$ from our model is currently $\sim \SI{25}{\gram\per\centi\meter\squared}$, which is worse than that obtained from the current approach using fluences, which yield a precision of $\leq \SI{19}{\gram\per\centi\meter\squared}$. We further notice that the values for $\log_{10}(N_\mathrm{max})$ are slightly underestimated. The origin of this bias is not yet known, and is under investigation.  \par 

\subsection{Conclusion}

To conclude, using an idealized star-shape antenna layout with a limited frequency bandwidth, our framework is able to reconstruct the entire longitudinal profile with uncertainties placed in each atmospheric bin. Furthermore, mass-sensitive parameters such as $X_\mathrm{max}$, $L$, and $R$ from the longitudinal profile can be reconstructed (with uncertainty) simultaneously, which not only significantly decreases the computational cost but can also strengthen current mass composition estimates \citep{Corstanje01.2026.SKA, Buitink01.2026.SKA}. \par 

We note that, while this approach is based on known parameterizations of physical models for near-field interferometric reconstruction of the air shower, model-agnostic approaches have also been explored. \cite{Straub:2025lsd} reconstructs the 3-D electromagnetic currents that generate the radio emission of the cosmic ray air shower at each 2-D atmospheric slice, where priors are introduced only through correlation between neighboring currents. With their approach, they are able to effectively reconstruct the full 3-D air shower profile in a model-agnostic fashion, using an idealized regular antenna grid. This approach, in complement with the parametric approach shown in this chapter, will be able to reveal more information about the intricate physics within the air shower, which can ultimately be used to better reconstruct the mass composition of cosmic rays.

\section{Outlook for SKA-Low AA$^*$}
\label{sec:Outlook}
We have showcased two methods that utilize Information Field Theory for a holistic reconstruction (Section~\ref{sec:HolisticRecoLOFAR}) and a near-field interferometric reconstruction (Section~\ref{sec:ProfileRecoTS}). While both approaches show great potential in terms of their applicability within the community, they have not yet been readily studied in the context of SKA-Low AA$^*$. In this section, we highlight how these approaches can benefit from the AA$^*$ antenna layout of SKA-Low, further highlighting its potential for next-generation cosmic ray measurements.

\subsection{Holistic Air Shower Reconstruction}
The results from Section \ref{sec:HolisticRecoLOFAR} already show the potential of a holistic air shower reconstruction on the example of LOFAR, yielding an $X_\text{max}$ precision of 25~\grammage, and a precision of \SI{12}{\percent} in radiation energy for a combined fluence-timing fit on a set of realistic simulations. As mentioned before, the LOFAR antennas are highly resonant, making reconstructing the full electric field spectrum challenging.  For the SKA-Low AA$^*$ we expect the antenna response to be much more uniform across its bandwidth. This will enable us to directly reconstruct on the electric field level as compared to first extracting fluence, timing and polarization data for the combined fit. It will also allow us to take direct information from the electric field spectrum, as its shape is also dependent on air shower parameters. This combined with the increased antenna density of the SKA-Low AA$^*$ layout will increase the precision of the method significantly, while keeping the computation time low compared to the simulation-based reconstruction approach. The method was already validated on a first set of LOFAR events \citep{terveer2026}, proving the feasibility of the method not only for simulations but for data as well.

\subsection{Air Shower Profile Reconstruction}

\begin{figure}
    \centering
    \includegraphics[width=0.48\linewidth]{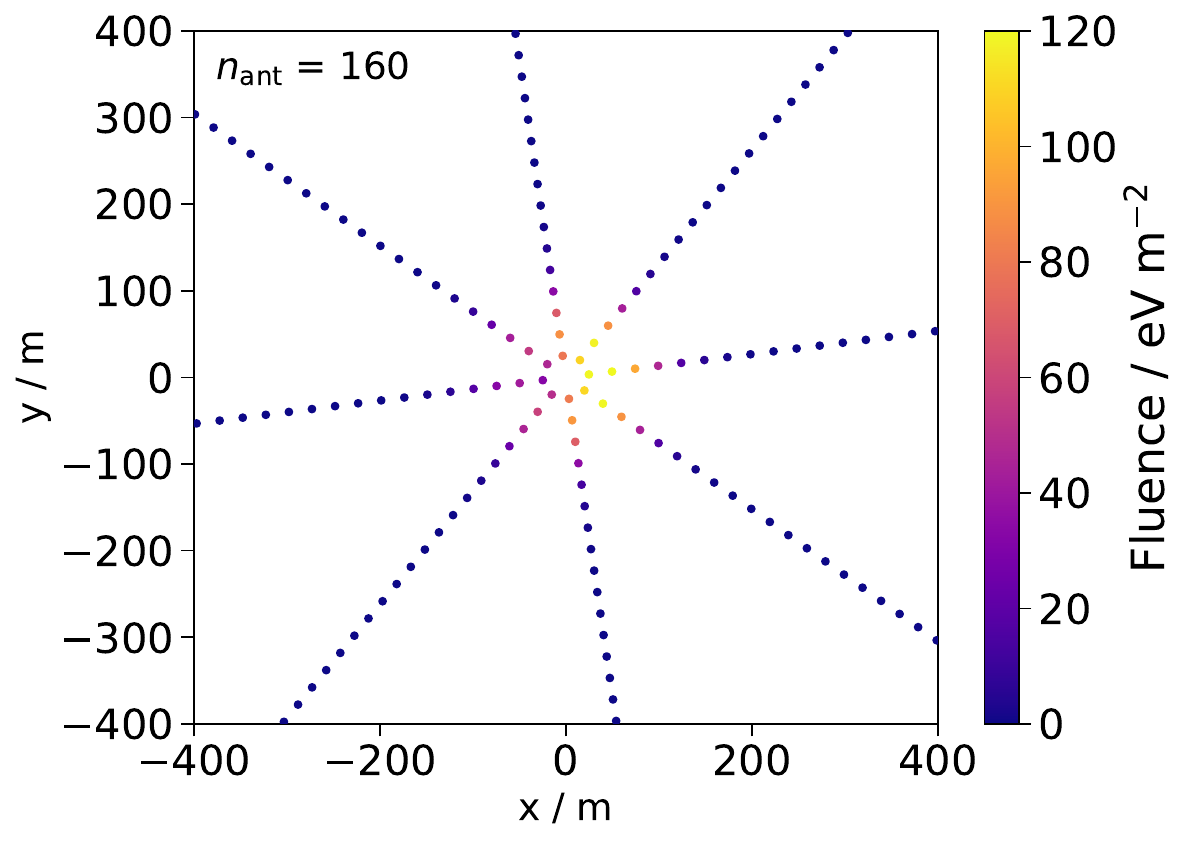}
    \includegraphics[width=0.48\linewidth]{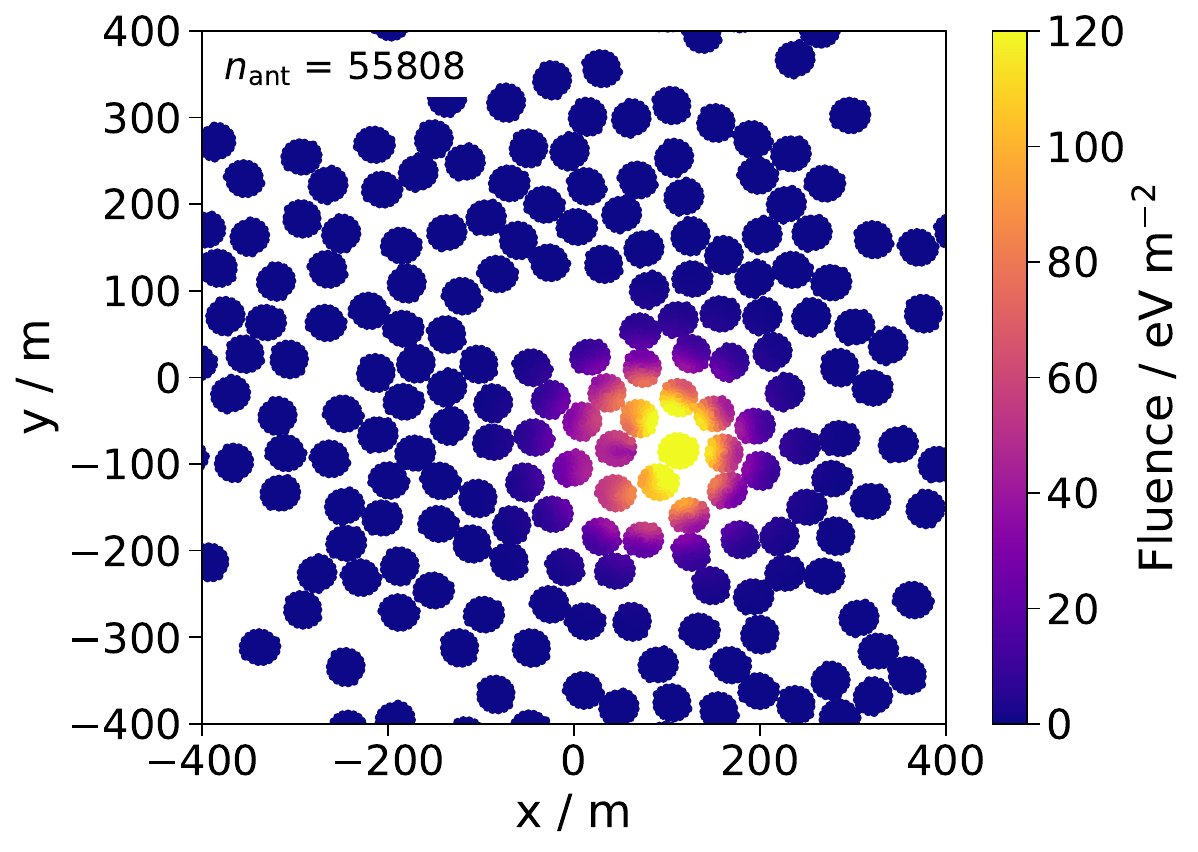}
    \caption{Left: Radio footprint (at ground) from a simulated air shower generated by a cosmic ray (with energy $E = \SI{3.8e17}{\electronvolt}$, zenith angle $\theta = 4^\circ$, and azimuthal angle $\phi = 151^\circ$), where the signal is observed in the idealized star-shape antenna layout. Right: Same simulated cosmic ray as the left figure, but for the SKA-Low AA$^*$ antenna layout. The shower core is shifted to (-100, 100) m to showcase an event that can be detected with maximal number of antennas. The total number of antennas for each configuration are also shown in the upper left. The frequency is bandwidth limited to [50, 350] MHz, emulating the proposed bandwidth for SKA-Low.}
    \label{fig:starshape_vs_ska}
\end{figure}

In Section~\ref{sec:ProfileRecoTS}, we showed that with an idealized star-shape antenna layout, we are able to reconstruct the entire longitudinal profile with a precision of reconstructing $X_\mathrm{max}$ of $\SI{25}{\gram\per\centi\meter\squared}$. Here, a total of 160 antennas were used in the analysis. As shown in Fig.~\ref{fig:starshape_vs_ska}, we can see that by using the SKA-Low AA$^*$ antenna layout, we not only increase the number of antennas used in the analysis (by more than a factor of 100), but also the antenna density within the same area, due to the small spacing between neighboring antennas. The increased antenna density and multiplicity has already been shown to be beneficial for cosmic ray measurements from \cite{corstanje_lofar-style_2025, Corstanje01.2026.SKA}, increasing the precision by a factor of 2 from those of LOFAR. As such we expect that using the SKA-Low AA$^*$ layout will also significantly increase the precision of our analysis. Furthermore, the near-field beamforming approach applied to SKA-Low AA$^*$ as shown in \cite{Corstanje01.2026.SKA} already highlights the capabilities for near-field interferometry. The larger frequency bandwidth and higher sampling rate are also crucial for better reconstruction, as more samples can be utilized within the reconstruction, which is essential to accurately reconstruct the pulse timing and shape of the cosmic ray signal. \par 

\section*{Acknowledgements}
The authors build on countless efforts to enable air shower observations using radio emission and are indebted to the community. Concretely, we acknowledge the following support: 
SBo, AN, and KT acknowledge the Verbundforschung of the German Ministry for Research, Technology and Space (BMFTR). 
PL and KW are supported by the Deutsche Forschungsgemeinschaft (DFG, German Research Foundation) – Projektnummer 531213488.
BH, CS, and PT are supported by ERC Grant Agreement No. 101041097. ST acknowledges funding from the Khalifa University RIG-S-2023-070 grant.
The authors gratefully acknowledge the computing time provided on the high-performance computer HoreKa by the National High-Performance Computing Center at KIT (NHR@KIT). This center is jointly supported by the Federal Ministry of Education and Research and the Ministry of Science, Research and the Arts of Baden-Württemberg, as part of the National High-Performance Computing (NHR) joint funding program. HoreKa is partly funded by the German Research Foundation. KM acknowledges funding from the Netherlands Research School for Astronomy (NOVA) and Dutch Research Council (NWO) project OCENW.XS25.1.237. This research is supported by the Flemish Foundation for Scientific Research (FWO-AL991 and FWO-OZR4291)

\clearpage

\bibliographystyle{abbrvnat}
\bibliography{chapter} 

\end{document}